\documentclass[amsmath,amssymb,prr,superscriptaddress,reprint,showpacs,longbibliography]{revtex4-1}
\usepackage{graphicx}
\usepackage{dcolumn}
\usepackage[colorlinks=true,linkcolor=blue,citecolor=blue,urlcolor=blue]{hyperref}
\usepackage[usenames,dvipsnames]{xcolor}
\usepackage{soul}
\usepackage{braket}
\usepackage{bm}
\usepackage{upgreek}
\usepackage{mathtools}
\usepackage{bbm}
\usepackage{multirow}

\renewcommand{\vec}[1]{{\bf #1}}

\begin{document}

\title{Twisted bilayers of thin film magnetic topological insulators}


\author{Gaurav Chaudhary}
\affiliation{Materials Science Division, Argonne National Laboratory, Lemont, IL 60439, USA}

\author{Anton A. Burkov}
\affiliation{Department of Physics and Astronomy, University of Waterloo, Waterloo, Ontario N2L 3G1, Canada}
\affiliation{Perimeter Institute for Theoretical Physics, Waterloo, Ontario N2L 2Y5, Canada}

\author{Olle G. Heinonen}
\altaffiliation{Present and permanent address: Seagate Technology, 7801 Computer Ave., Bloomington, MN55435}
\affiliation{Materials Science Division, Argonne National Laboratory, Lemont, IL 60439, USA}

\date{\today}



\begin{abstract}
Twisted bilayer graphene (TBG) near ``magic angles'' has emerged as a rich platform for strongly  correlated states of two-dimensional Dirac semimetals. Here we show that twisted bilayers of thin-film magnetic topological insulators (MTI) with large in-plane magnetization can realize flat bands near 2D Dirac nodes. 
Using a simple model for thin films of MTIs, we derive a continuum model for two such MTIs, twisted by a small angle with respect to each other. 
When the magnetization is in-plane, we show that interlayer tunneling terms act as effective $SU(2)$ vector potentials, which are known to lead to flat bands in TBG.
We show that by changing the in-plane magnetization, it is possible to tune the twisted bilayer MTI band dispersion to quadratic band touching or to flat bands, similar to the TBG. 
If realized, this system can be a highly tunable platform for strongly correlated phases of two-dimensional Dirac semimetals.
\end{abstract}

\maketitle

\section{Introduction}
The discovery of correlated insulators~\cite{Cao2018a} and superconductivity~\cite{Cao2018} in the magic angle twisted bilayer graphene (MATBG) has given birth to a new paradigm in the physics of strongly correlated electron systems. 
When the two graphene layers are twisted with respect to each other, the Dirac dispersive bands of the individual layers are strongly renormalized and at certain twist angles become extremely flat~\cite{Santos2007,Bistritzer2011}.   
In these flat bands electron-electron interactions may dominate and give rise to interesting strongly correlated phases~\cite{Sharpe2019,Serlin2019,Kerelsky2019,Xie2019,Jiang2019}. The twist angle then becomes an important tuning knob to realize such phases. 
Moreover, the Dirac dispersion of the underlying graphene layers adds topological character to the flat bands~\cite{Bouhon2019,Ahn2019,Song2019,Xie2019a,Lu2021}. 
Hence, this system is potentially a fertile ground for the interplay of topology and 
correlations, which often leads to interesting and exotic phases~\cite{Serlin2019,Lu2019}. 

Inspired by these findings, twist engineering of stacked two-dimensional layered materials has been explored in a variety of other systems, such as transition metal dichalcogenides~\cite{Wu2018,Wu2019,Tang2020,Regan2020,Xu2020,Zhang2020,Slagle2020,Bi2021,Duran2021}, topological insulator (TI) surface states~\cite{Lian2020,Liu2021,Cano2021,Wang2021,Dunbrack2021}, and cuprate superconductors~\cite{Volkov2020,Can2021,Zhao2021}, to name a few. 
Among these, the TI surface states potentially have closest resemblance to the TBG case due to Dirac crossing bands structures similar to graphene. 
However, twist engineering flat bands like MATBG has been shown to be difficult for the TI surface states~\cite{Cano2021,Dunbrack2021}. 

In this work, we consider twisted bilayers of thin film magnetic topological insulators (MTI). 
When a single layer of an MTI has large in-plane magnetization, it develops two isolated two-dimensional (2D) Dirac nodes, separated in the momentum space~\cite{Burkov2018}. 
Using these Dirac nodes as building blocks, we construct a simple continuum model for twisted bilayer magnetic topological insulators (TBMTI). 
We show that in this setup, one can overcome many of the difficulties of twist engineering flat bands in TI surface states. 
We demonstrate that depending on the momentum position of the Dirac dispersion of an isolated MTI, one can engineer quadratic band touching (QBT) points or extremely flat bands in the TBMTI. 
We attribute the appearance of the flat bands to emergent $SU(2)$ vector potentials~\cite{Wilczek1984}, that depend on interlayer tunneling and magnetization. 
In the small twist angle limit, these emergent vector potential lead to local zeroth pseudo Landau levels (pLL), which are shown to be responsible for robust flat bands. 
We also show that the magnetization plays an important role in tuning the band structures; both because of its role in breaking mirror symmetries, protecting the Dirac nodes, and in determining the strength of the effective interlayer tunneling. 
Thus the twist control in TBG can be mimicked in this system by simply tuning the magnetization at a fixed twist angle. 
If realized, this system can be a highly tunable platform for strongly correlated phases of 2D Dirac semimetals. 
These potential correlated phases may be fundamentally different from the correlated phases of 2D Dirac semimetals realized in MATBG, because of the absence of spin degeneracy due to time reversal $(\mathcal{T})$ symmetry breaking.

\begin{figure}[t]
  \includegraphics[width=0.5\textwidth]{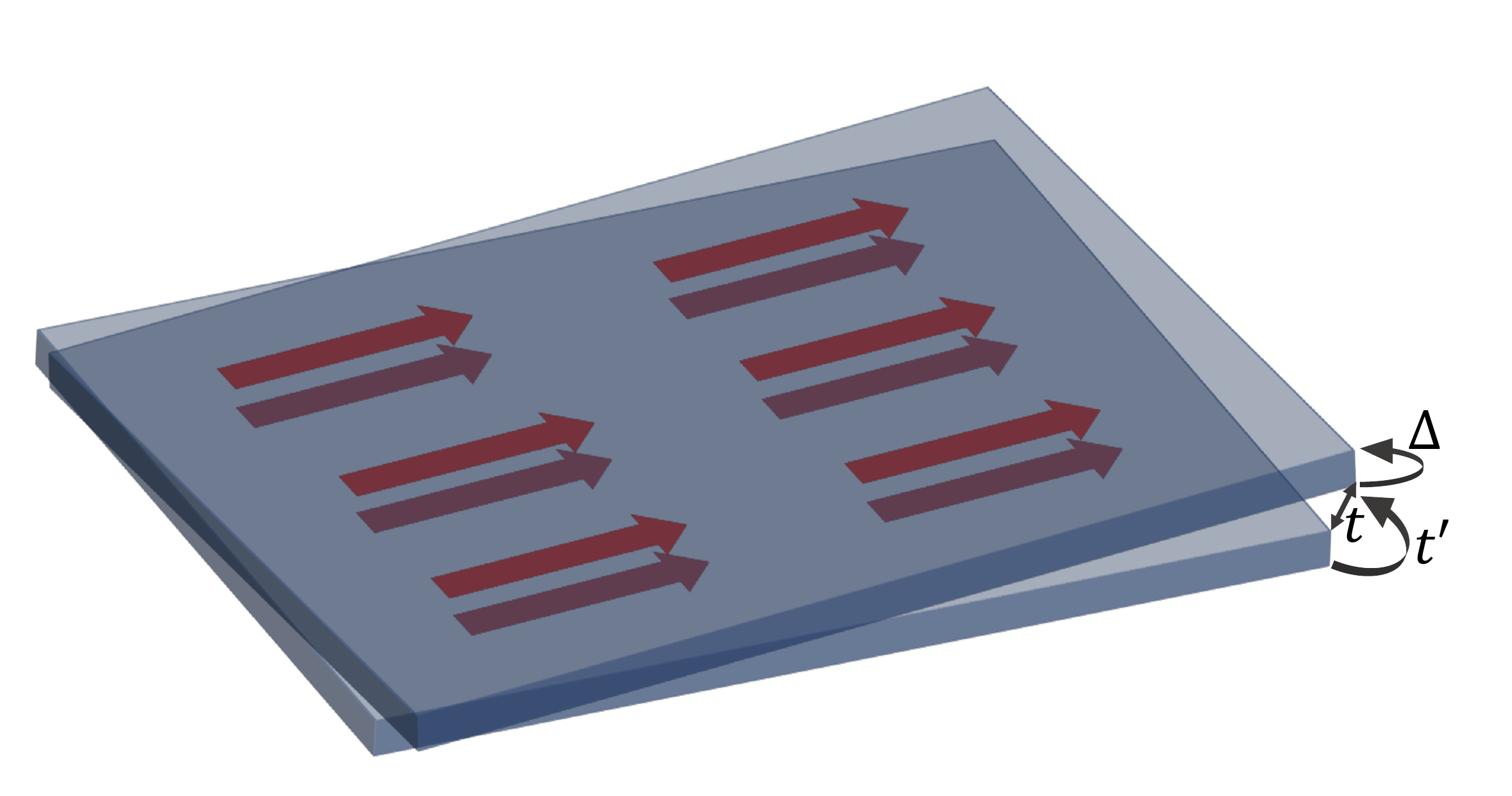}
  \caption{\label{Fig:TBMTI}
  Twisted bilayers of thin film magnetic topological insulators. The red arrows represent in-plane magnetization. Electrons can tunnel between surfaces within one layer, described by $\Delta$, or between nearest (next-next nearest) surfaces of the two different layer, described by $t$ ($t'$), with    
  $|t|>|\Delta|>|t'|$.
}
\end{figure}

\section{Setup and Model} 
We start with a minimal description of MTI thin films, using a coupled model of top and bottom surface Dirac cones with an explicit magnetization term that breaks $\mathcal{T}$-symmetry. 
If the magnetization exceeds the inter-layer tunneling and has in-plane orientation, perpendicular to a mirror plane, the system develops two Dirac nodes in the low energy sector~\cite{Burkov2018}. 
With an aim to derive a continuum model of TBMTI near these Dirac nodes, we start with a square-lattice model for the surface states of a single thin film of an MTI. In the momentum space, the surface-spin basis Hamiltonian takes the form:
\begin{align}\label{Eq:Ham_MTI}
     & H = v (\sin k_y \sigma_x - \sin k_x \sigma_y)\otimes \tau_z + \vec{m}\cdot\bm{\sigma}  + d_{\vec{s}}(\vec{k}) \tau_x ,
\end{align} 
where, $v$ is the Fermi velocity of the surface Dirac dispersions, $\vec{m}$ is the  magnetization, $\sigma$ and $\tau$ Pauli matrices act on spin and surface respectively, and  
\begin{align}\label{Eq:Ham_inter_surface}
    d_{\vec{s}}(\vec{k}) =  \Delta + \Delta' (2 + s_x \cos k_x + s_y \cos k_y )
\end{align}
contributes to the inter-surface Hamiltonian. Microscopically, $\Delta$ and $\Delta'$ can be interpreted as nearest neighbor and next-nearest neighbor inter-surface tunnelings within one layer. 
Notice that $\Delta'$ breaks the degeneracy between the time-reversal invariant momenta (TRIM) ($\Gamma$, $X$, $Y$, and $M$ points). 
The label $\vec{s} = (s_x,\, s_y)$, where $s_x$ and $s_y$ take values $\pm 1$ to determine the location of the surface Dirac node (as described below). 
We have chosen $\hbar = a = 1$. In lattice constant units, the reciprocal unit cell vectors are $\vec{g}_1 = (2\pi,\,0)$ and $\vec{g}_2 = (0,\,2\pi)$. 

The four energy bands of the Hamiltonian in Eq.~\ref{Eq:Ham_inter_surface} are
\begin{align}\label{Eq:Spectrum_MTI}
    & E(\vec{k}) = \pm \biggl ( d_{\vec{s}}^2 + m^2 + v^2\sin^2 k_x + v^2\sin^2 k_y \notag\\
    &\hspace{1cm} \pm 2 \sqrt{d_{\vec{s}}^2 m^2 + v^2(m_x \sin k_y - m_y \sin k_x)^2 }   \biggr )^{1/2}.
\end{align}
In the limit, $\vec{m}\rightarrow \vec{0}$ and  $\Delta\rightarrow 0$, the surface Dirac nodes remain protected. 
If $s_x=s_y=-1$, the Dirac node is at the zone center ($\Gamma$-point) and if $s_x=s_y=1$, the Dirac node is at the zone corner ($M$-point). 
These two cases are invariant under the $C_{4z}$ rotation and mirror reflections in four perpendicular planes of the square lattice ($xz$, $yz$, and their bisector planes).  
The $s_x \neq s_y$ cases are less symmetric because the $C_4$ symmetry of the square lattice and two out of four mirror symmetries are broken. 
When $s_x (s_y) = 1, s_y (s_x) = -1$, the Dirac point appears at $X (Y)$-point. 
A finite value of $\Delta$ tends to gap out the Dirac node in a trivial fashion (semimetal-to-trivial insulator transition) and a finite value of out-of-plane magnetization component tends to gap out the Dirac node in a topological fashion (semimetal-to-Chern insulator transition). 
Thus as a function of $\Delta$ and $\vec{m}$ the system undergoes a Chern insulator-trivial insulator phase transitions. 

\begin{figure*}[t]
  \includegraphics[width=1\textwidth]{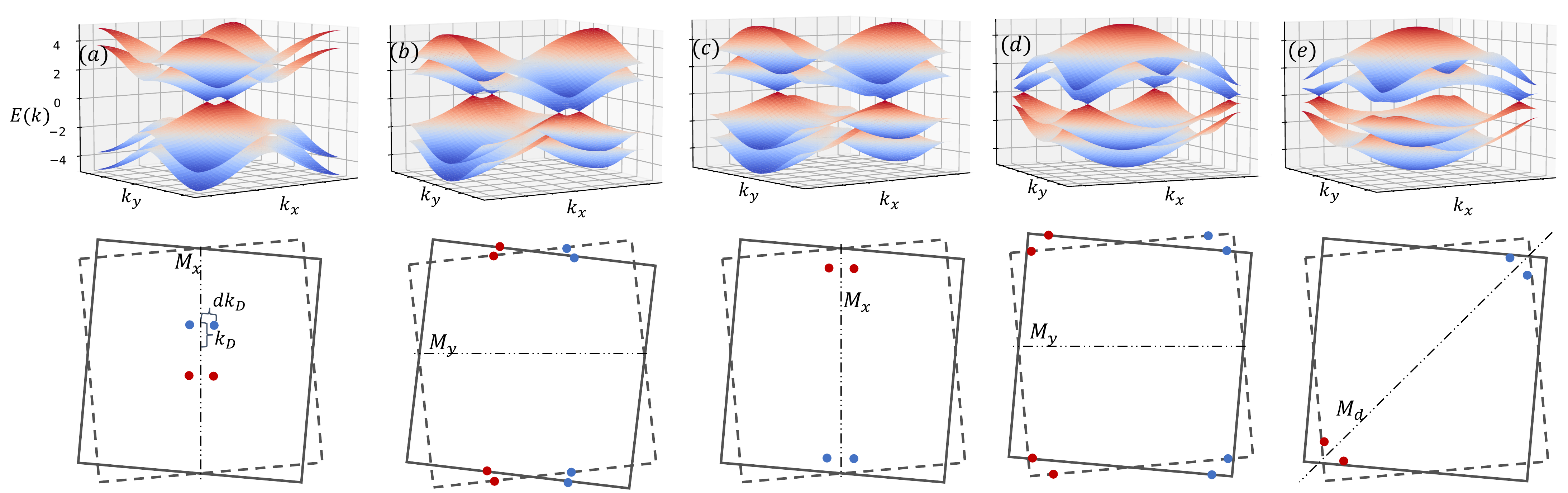}
  \caption{\label{Fig:MTI_model_schematic}
  Band structure and Dirac nodes for in plane magnetization: (a) $s_x = s_y = -1$, $\vec{m} = (m,0,0)$, (b) $s_x = -s_y = 1$, $\vec{m} = (0,m,0)$, (c) $s_x = -s_y = 1$, $\vec{m} = (m,0,0)$, (d) $s_x=s_y=1$, $\vec{m}=(0,m,0)$, and (e) $s_x = s_y = 1$, $\vec{m} = \frac{1}{\sqrt{2}}(-m,m,0)$. The bottom panel shows the respective twisted bilayer configuration with location of Dirac nodes and relevant mirror planes.
}
\end{figure*}

An interesting situation occurs when the magnetization is perpendicular to a mirror plane~\cite{Burkov2018}. As the magnetization is increased, the gap closes at the location of the original surface Dirac point at a critical value $m = \Delta$. 
Further increase in magnetization leads to a pair of Dirac points, separated in momentum space by $\delta k_D = 2\sqrt{m^2-\Delta^2}/v$ in the direction perpendicular to the magnetization. 
As a concrete example, consider the case when $s_x = s_y = -1$ in the limit of linear  expansion in momentum near $\Gamma$-point, and magnetization along the $x$-direction [See Fig.~\ref{Fig:MTI_model_schematic} (a)]. The spectrum along $k_x = 0$ line then reduces to
$E(k_y) = \pm (m \pm \sqrt{\Delta^2 + v^2 k^2_y} \,)$, with Dirac crossings at $\pm \vec{k}_D = (0,\, \pm \sqrt{m^2-\Delta^2}/v)$. 
These Dirac nodes are protected by mirror symmetry $M_x$ (mirror $yz$ plane). 
In our analysis, such Dirac nodes will serve as the analogue of the Dirac nodes of opposite valleys in graphene and will be the building block of our continuum model for TBMTI. 

We end this section with some comments on the symmetries of $H$.
When the magnetization is zero, the system has $\mathcal{T}$-symmetry, inversion ($\mathcal{I}$)-symmetry and a set of symmetries related to the square lattice. 
In case of $s_x = s_y$, these square lattice symmetries are, the $C_{4z}$ rotation, mirror reflections $M_x$, $M_y$, mirror reflections ($M_d$) in the two diagonal planes bisecting $xz$ and $yz$ planes, $C_{2x}$, $C_{2y}$ rotations, and two $C_{2d}$ rotations around the two diagonal in-plane axes.  
An in-plane magnetization, along with the $\mathcal{T}$-reversal, also breaks some of the mirror and rotation symmetries. 
We are interested in scenarios, when the magnetization is perpendicular to one of the  mirror planes. 
It turns out that only the mirror perpendicular to the magnetization, $C_2$ rotation around the magnetization axis, and $\mathcal{I}$ survive, while $C_{4z}$ rotation and $\mathcal{T}$ are broken to a reduced  $C_{2z}\mathcal{T}$ symmetry. 
The $C_{2z}\mathcal{T}$ exchanges the two Dirac nodes in the low energy sector (See the bottom panels of Fig.~\ref{Fig:MTI_model_schematic}).

\subsection{Bilayer magnetic topological insulator}
Having discussed the single layer model for thin film MTI, we now move on to bilayers. 
The untwisted bilayer can be described by a simple Hamiltonian:
\begin{align}\label{Eq:Ham_bilayer}
    H_{BL} = \begin{pmatrix}
    H_{t} & T \\
    T^{\dagger} & H_b \\
    \end{pmatrix} ,
\end{align}
where, $H_t,H_b \equiv H$ describe the top and bottom MTI layers. 
The tunneling Hamiltonian 
\begin{align}\label{Eq:Ham_tunneling}
    T = t' \sigma_0\otimes \eta_0 + \frac{t}{2}\sigma_0\otimes (\eta_x + i \eta_y)
\end{align} 
only has spin-preserving tunneling processes and $\eta_i$ are Pauli matrices. 
Physically, $t'$ is the tunneling between top(bottom) surface-to-top(bottom) surface tunneling between two MTIs and $t$ is the tunneling between top surface of the bottom layer to the bottom surface of the top layer. 
Thus $|t| > |\Delta | > |t'| $, and they respectively describe tunnelings between nearest (inter-layer), next nearest (intra-layer), and third nearest (inter-layer) surfaces. 

Recently, it was shown that in twisted topological insulators, the spin-preserving tunnelings act as an $SU(2)$ scalar potential~\cite{Dunbrack2021}. 
In TBG, an analogous scalar potential term originates from the intra-sublattice tunnelings~\cite{San-Jose2012}, while the inter-sublattice tunneling acts as an $SU(2)$ vector potential, responsible for the flat bands~\cite{San-Jose2012,Tarnopolsky2019}.  
Based on this, it was concluded that one requires spin-flip tunnelings to obtain flat bands in twisted topological insulators~\cite{Dunbrack2021}, since they are analogous to the inter-sublattice tunneling in TBG. 
 However, spin-flip tunneling amplitudes involve overlap between the up and down-spin 
components of the surface state wavefunctions, which vanishes upon momentum integration in the vicinity of a Dirac point.
Thus the twisted topological insulators are expected to have vanishing effective $SU(2)$ vector potentials. 
Here, we show that this limitation can be overcome when $\mathcal{T}$-symmetry breaking magnetization has finite in-plane component. 

For this purpose, we expand the MTI Hamiltonian to linear order in momentum near a TRIM $\vec{k}_0$ (center of the Dirac dispersion of the isolated layer) and then transform it to a form similar to the standard representation of Dirac Hamiltonian. 
Next, we assume that the two Dirac nodes in a single MTI are decoupled and can be considered as two independent valleys, such that the linearly expanded system near the Dirac nodes is approximated as $H \sim H_{+}(\vec{k}_{D+}+d\vec{k}) \oplus H_{-}(\vec{k}_{D-}+d\vec{k})$. 
In this limit, the Hamiltonian near one of the valley (Dirac node), after another similarity transformation can be recast in the form (see the App.~\ref{Sec:App_bilayer_derivation} for the derivation):
\begin{align}\label{Eq:Ham_MTI_transformed}
    H_{\chi}(\vec{k}) = \begin{pmatrix}
    \chi v_\perp k_y \gamma_x + v k_x \gamma_y & 0_{2\times 2} \\
    0_{2\times 2} & (2m + \chi v_\perp k_y) \gamma_x - v k_x \gamma_y
    \end{pmatrix},
\end{align}
where, $v_\perp = [1-\Delta^2/(2m^2)]v$, and $k_x,\, k_y$ are measured from the Dirac node of ``chirality" $\chi$. 
Here, $\gamma_i$ are Pauli matrices. 
From here on, we use $\eta_i$ and $\gamma_i$ Pauli matrices for unspecified bases to give a $2\times 2$ matrix structure to our equations where required. 
In obtaining Eq.~\ref{Eq:Ham_MTI_transformed}, we have also performed an in-plane axes rotation if required. 
For example, if the nodes are near the Brillouine zone (BZ) corner as shown in Fig.~\ref{Fig:MTI_model_schematic} (d), (e), we perform a $\pi/4$-clockwise axes rotation to obtain Eq.~\ref{Eq:Ham_MTI_transformed}. 

In $H_{\chi}(\vec{k})$, the upper-diagonal block describes the low energy sector that contains the Dirac crossings, while the lower-diagonal block describes a gapped high energy sector. 
The transformed interlayer Hamiltonian along with coupling the low (high) energy sector of two layers, also couples the low energy sector from one layer to the high energy sector in the other layer. 
However, because of the energy gap between these two sectors, this coupling only leads to small quantitative corrections. 
Thus it suffices to restrict to the low energy sector that contains the protected Dirac crossings. 
In the transformed basis the interlayer tunneling in the low energy sector takes the form
\begin{align}\label{Eq:Ham_tunnel_transformed}
    T = \begin{pmatrix}
        t' & -\frac{t\Delta}{2m} \\
        -\frac{t\Delta}{2m} & t'
    \end{pmatrix} . 
\end{align}
Since, $t'$ is the third nearest neighbor tunneling between surface states, for most of our analysis, we approximate $t'\sim 0$. 
The final low energy bilayer Hamiltonian near a Dirac node $\chi$ can be approximated as 
\begin{align}\label{Eq:Ham_BL_approx}
    H_{\chi, BL} = \begin{pmatrix}
        H_{\chi,l} (\vec{k}) & t'\eta_0-\frac{t\Delta}{2m} \eta_x \\
        t'\eta_0-\frac{t\Delta}{2m} \eta_x & H_{\chi,l} (\vec{k})
    \end{pmatrix}. 
\end{align}
Here $H_{\chi,l}$ is the low energy sector given by the upper-diagonal block in Eq.~\ref{Eq:Ham_MTI_transformed}.
The most important observation from the above form of the bilayer Hamiltonian near the Dirac node is the  appearance of off-diagonal terms in tunneling, which will appear as an $SU(2)$ vector potential when the bilayers are twisted. 
When $t'= 0$, the  Eq.~\ref{Eq:Ham_BL_approx} has a chiral symmetry and we expect that twisted bilayers will closely resemble the chiral symmetric model of the TBG~\cite{Tarnopolsky2019}. 
However, the Dirac nodes here are anisotropic, and unlike TBG, are not pinned to high symmetry points  in the moir\'e Brillouin Zone (mBZ), which will lead to  some qualitative differences. 
Since we ignore coupling between opposite chirality nodes, we assign $\chi = +$ to the nodes depicted in blue in the bottom panels of Fig.~\ref{Fig:MTI_model_schematic} and only consider these nodes in all our calculations.

\subsection{Twisted bilayer magnetic topological insulators}
When the two MTI layers are twisted with respect to each other by a small angle $\theta$, the interlayer tunneling becomes spatially dependent because of the long-range moir\'e potential. 
It is sometimes more convenient to represent the Hamiltonian in real space. 
Thus, we replace the momenta by appropriate differential operators and derive the continuum model applicable in the low-energy sector near the Dirac nodes. 
The continuum model for TBMTI in the spirit of the Bistritzer-MacDonald (BM) model for TBG~\cite{Bistritzer2011} takes the form (see the App.~\ref{Sec:App_TB_derivation} for the derivation):
\begin{align}\label{Eq:Ham_TBMTI}
    &H_{TB} = \int d^2\vec{x} \{ \psi^{\dagger}_t[-i\bm{\gamma}(\theta/2)\cdot \bar{\bm{\nabla}} + \frac{1}{2}(\vec{k}_0+\vec{k}_D)\cdot\bar{\partial}_\mu \vec{u}  \gamma_{\mu} ] \psi_t \notag\\
     &\hspace{1.5cm}+ \psi^{\dagger}_b[-i\bm{\gamma}(-\theta/2)\cdot \bar{\bm{\nabla}} - \frac{1}{2}(\vec{k}_0+\vec{k}_D)\cdot\bar{\partial}_\mu \vec{u} \gamma_{\mu} ] \psi_b \notag\\
     &\hspace{2.5cm}+ [  \psi^{\dagger}_{t} T(\vec{x}) \psi_{b} + \text{h.c.} ] \},
\end{align} 
where $\bm{\gamma} (\theta)  = (\text{e}^{-i\frac{\theta\gamma_z}{2r}}\gamma_y \text{e}^{i\frac{\theta\gamma_z}{2r}}, \text{e}^{-i\frac{ r \theta  \gamma_z}{2}}\gamma_x \text{e}^{i\frac{r \theta  \gamma_z}{2}}) $, $r = v_x/v_y$,  $\psi_t (\psi_b)$ is the top (bottom) two-component Dirac electron annihilation operator, $\vec{k}_D$ is the momentum space position of the `$\chi=+$' Dirac node before twist measured from a TRIM $\vec{k}_0$, 
\begin{align}\label{Eq:tunneling_expression}
    &T (\vec{x}) = \sum_{n,m\in \mathbb{Z}}\text{e}^{-i\vec{G}_{nm} \cdot \vec{u} (\vec{x})}  (t_{nm,0}\eta_0+t_{nm,x}\eta_x ), 
\end{align}
such that $t_{nm,i} = t^{\ast}_{-nm,i}$ is the $M_x\eta_x$ (where $\eta_x$ exchanges top and bottom layers) symmetric interlayer tunneling Hamiltonian that has spatial dependence due to moir\'e modulation (see App.~\ref{Sec:App_TB_derivation} for derivation), $\vec{G}_{n,m} = n\vec{g}_1+m \vec{g}_2$, and $\vec{u} = \theta\hat{z}\times \vec{x} $ is relative deformation field between the layers associated with the rigid twist. 
Comparing Eq.~\ref{Eq:tunneling_expression} and Eq.~\ref{Eq:Ham_tunnel_transformed}, $t_{nm,0}$ and $t_{nm,x}$ are the Fourier components of the moir\'e modulated diagonal and off-diagonal tunneling terms respectively. 
The renormalized operators $\bar{\bm{\nabla}} = (\bar{\partial_x},\,\bar{\partial_y}) = (v_x\partial_x,\, v_y\partial_y)$, take into account the anisotropic Fermi velocity of the nodes. 
The above Hamiltonian has as $\vec{k}_0+\vec{k}_D$ as its origin. 
Notice that in writing $H_{TB}$ as above, we have performed an appropriate rotation around $z$-axis (if required depending on the $\vec{k}$-space position of the Dirac nodes). 
It should be kept in mind that any such an axis rotation should also be accounted for in rotation of the reciprocal lattice vectors $\vec{g}_i$.
After this rotation, $M_x\eta_x$ is the only relevant symmetry for all the cases. 
It should be noted that $C_{2x}$ is another symmetry that acts identically to $M_x\eta_x$ in the low energy projected space. 
Finally, we mention that Hamiltonian near $\chi = -$ nodes can be obtained by $C_{2z}\mathcal{T}$ symmetry.

We can remove the Dirac node shift terms in the individual layer by gauge transformation $\bar{\psi}_{t/b} = \text{e}^{\pm i r \theta \gamma_z/4} \text{e}^{\mp i\theta \hat{z}\times(\vec{k}_0+\vec{k}_D)\cdot \vec{x}/2}\psi_{t/b}$ and writing $\bar{\psi} = (\bar{\psi}_t,\, \bar{\psi}_b)$, the Hamiltonian can be expressed as a 2D Dirac Fermion in $SU(2)$ gauge potentials,
\begin{align}\label{Eq:Ham_TBMTI_gauge_form}
    H_{TB} \sim \int d^2\vec{x} \bar{\psi}^{\dagger} [ \gamma_x (-i\bar{\partial}_y - \bar{T}_x ) 
    -i \gamma_y\bar{\partial}_x + \bar{T}_0 \gamma_0 ]\bar{\psi}.
\end{align} 
Here
\begin{subequations}
\begin{align}
    & \bar{T}_{0}(\vec{x}) = \text{e}^{i\frac{r\theta}{2}\gamma_z}\sum_{n,m} \text{e}^{i(\vec{k}_0+\vec{k}_D-\vec{G}_{nm})\cdot \vec{u}(\vec{x})} t_{nm,0}\eta_{0}\, \\
    & \bar{T}_{x}(\vec{x}) = \sum_{n,m} \text{e}^{i(\vec{k}_0+\vec{k}_D-\vec{G}_{nm})\cdot \vec{u}(\vec{x})} t_{nm,x}\eta_{x}.
\end{align} 
\end{subequations}

In writing Eq.~\ref{Eq:Ham_TBMTI_gauge_form}, we have made the small angle approximation and ignored a term $\sim \Delta^2\theta/(4m^2)\gamma_x\eta_z$, owning to its very small magnitude. 
Since $\bar{T}_x$ acts as a Peierls shift in the $y$-momentum, it is the $y$-component of a vector potential. In our choice, the $x$-component of the vector potential is zero, since there are no $\eta_y$ terms in tunneling. 
Similarly, $\bar{T}_0$ is the scalar potential. 
These $SU(2)$ potentials are periodic in space with vanishing spatial average.
Because $\mathcal{T}$-symmetry is broken $t_{nm,i}$ can take complex values. 
In our calculation, we only restrict to their real values for simplicity. 
Notice, $T(\vec{x})$ is periodic under moir\'e unit vector translation. 
However, since Dirac nodes are not pinned to a high symmetry point, $\bar{T}(\vec{x})$ breaks the moir\'e translation symmetry. 
This is an artefact of continuum model near the Dirac node.

\section{Flat bands and their origin}

We calculate the resultant low energy moir\'e band of TBMTI by diagonalizing Eq.~\ref{Eq:Ham_TBMTI}, with a suitable cutoff on tunneling Fourier components. 
In the real systems, the inter-surface distance is much larger than the near neighbor  inter-site distance within a surface. 
As a consequence, the interlayer tunneling decays slowly at the scale of the microscopic lattice constant and thus rapidly in momentum space relative to the reciprocal lattice vector $\vec{G}_{nm}$.  
Thus it suffices to keep only a few low order Fourier components. 
This is a standard approximation used in the BM model for TBG~\cite{Bistritzer2011}. 
In our case, the smallest Fourier components are determined by lowest possible magnitudes of $\vec{G}_{n,m}-\vec{k}_0$, for the TRIM point $\vec{k}_0$. 
Thus we analyze different cases of $\vec{k}_0$ separately.

\subsection{$\Gamma$-point}
When the Dirac dispersion of the isolated layer is centered around the $\Gamma$-point, the lowest possible Fourier components are $n = m = 0$, which also predominantly contribute to the interlayer tunneling. 
In the approximation where only $\vec{G}_{00}$ component is kept, the interlayer tunneling has no real space moir\'e modulation. 
Assuming $t'\sim 0$, we obtain an analytic expression for the band dispersions centered around $\vec{k}_D$:
\begin{widetext}
\begin{align}\label{Eq:TBMTI_Gamma_energy}
    & E(\vec{k})= \pm \sqrt{v^2_x k^2_x \biggl (1-\frac{k^2_D}{t^2_{00,x}}\sin^2\frac{\theta}{2} \biggr )  + \biggl ( t_{00,x} \pm \sqrt{v^2_y \biggl [ k_y- 2k_D \sin^2\frac{\theta}{4} \biggr ]^2 + v^2_x k^2_D\sin^2 \frac{\theta}{2} \biggl [ 1 +\frac{v^2_x k^2_x}{t^2_{00,x}} \biggr ]
    } 
    \biggr )^2
    }.
\end{align}
\end{widetext}
Here $v_x = v$, $v_y = v(1-\Delta^2/(2m^2))$, and $t_{00,x} = - t\Delta/(2m)$. 
This dispersion has gapless points originating from the Dirac nodes of the two layers that, we find remain protected because of the $M_x\eta_x$-symmetry. 
Interestingly at twist angle
\begin{align}\label{Eq:magic_angle_condition}
    \theta_M = 2\arcsin \biggl ( \frac{t\Delta}{2m\sqrt{m^2-\Delta^2}}\biggr )
\end{align}
the two same chirality Dirac nodes of the two layers merge to form a QBT point at $\vec{k}_D$.

\begin{figure}[h]
  \includegraphics[width=0.48\textwidth]{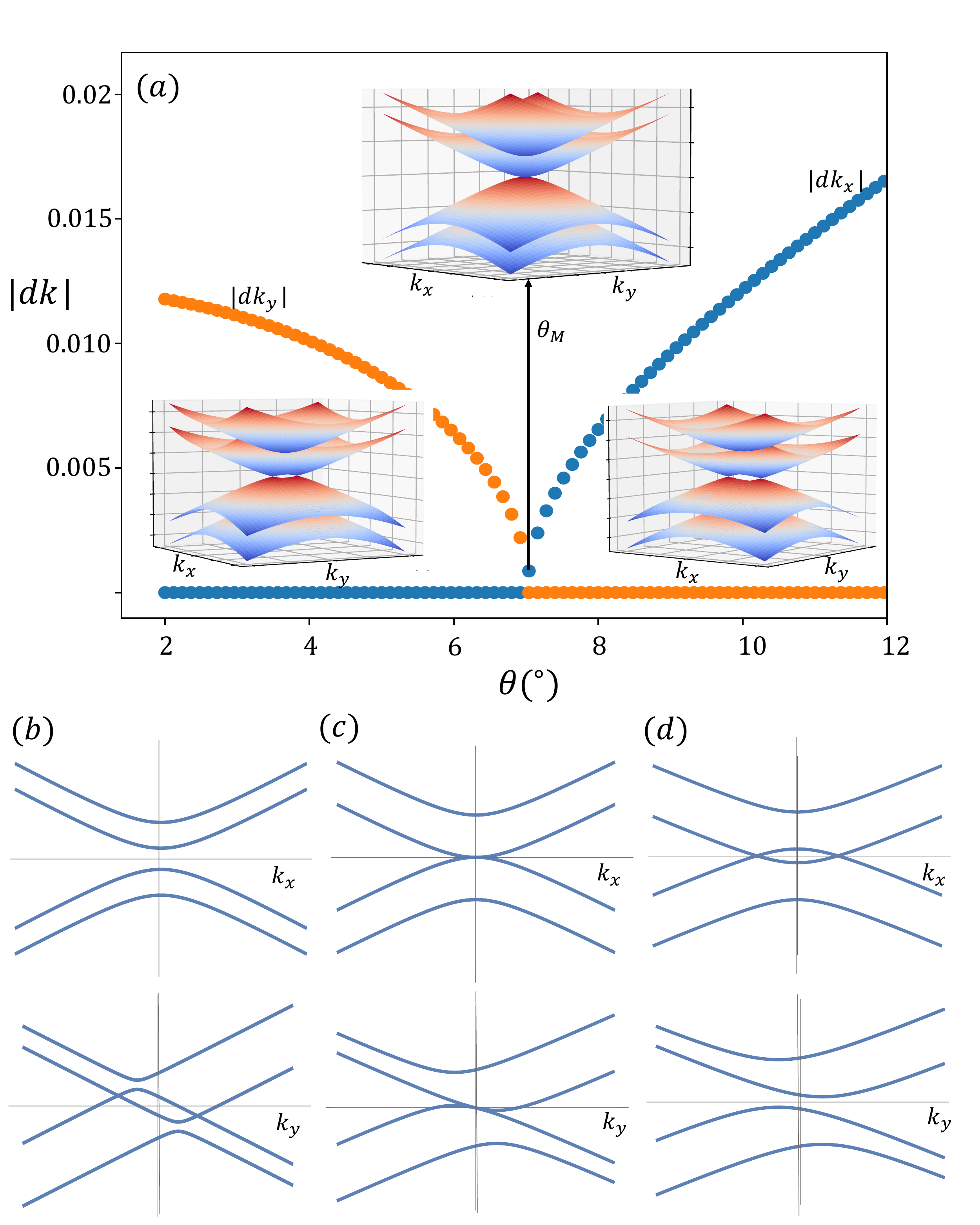}
  \caption{\label{Fig:Gamma_QBT}
  Quadratic band touching in TBMTI: (a) Separation between Dirac nodes from top and bottom layer of MTI as function of twist angle. The blue (orange) markers represent the separation along $k_y (k_y)$ direction. The two Dirac nodes merge at $\theta_M$ to form QBT point. (b)-(d) The bands structure when third nearest surface tunneling ($t'$) is included for $\theta > \theta_M$, $\theta = \theta_M$, and $\theta > \theta_M$. The top panels are plotted along $k_x$ at $k_y = 0$, the bottom panels are plotted along $k_y$ at $k_x = 0$. 
  The above plots are obtained for $v = 1$ eV$\cdot\AA$, $m = 100$ meV, $\Delta = 20$ meV, $t = 60$ meV, and $t' = 4$ meV [(b)-(d)].
}
\end{figure}

The emergence of the QBT at an isolated twist angle becomes more transparent by studying the trajectory of the Dirac nodes from the two layers as shown in the Fig.~\ref{Fig:Gamma_QBT} (a). 
In the decoupled layer limit, the two `$\chi = +$' Dirac nodes of the two layers are at $\vec{k} = \vec{k}_D$. 
As the interlayer tunneling is turned on, the Dirac nodes repel each other and for untwisted configuration separate along the mirror plane ($y$-direction). 
If we start twisting the two layers with respect to  each other, the Dirac nodes first move along the mirror plane towards each other and merge at $\theta_M$ to form QBT point as shown in the left half of the Fig.~\ref{Fig:Gamma_QBT} (a). 
As twist angle is increased further, the nodes separate again by moving along direction perpendicular to the mirror plane as shown in the right half of the Fig.~\ref{Fig:Gamma_QBT} (a). 
In Fig.~\ref{Fig:Gamma_QBT} (b)-(d), we show the band dispersions when a finite $t_{00,0}\equiv t'$ tunneling is added. This term breaks the chiral symmetry [bottom panels of Fig.~\ref{Fig:Gamma_QBT} (b)-(d)] along $k_y$. 
For $\theta < \theta_M$, the Dirac nodes separate in energy to form type-II nodes [the bottom panel in Fig.~\ref{Fig:Gamma_QBT}(b)]. 
We again find a QBT [see Fig.~\ref{Fig:Gamma_QBT} (c)], when the two Dirac nodes meet as the twist angle is tuned.
Moreover, at QBT, the system has electron-hole Fermi surfaces. 
Finally for $\theta > \theta_M$, the system recovers Dirac nodes at zero energy separated along $k_x$. 
We mention that for the finite $t'$ case, $\theta_M$ is not given by the simple expression of Eq.~\ref{Eq:magic_angle_condition}.

The phenomena described above has been recently proposed for the Bogoliubov quasiparticle bands in twisted bilayers of cuprate superconductor and the QBT angle was called the ``magic angle" in this case~\cite{Volkov2020}. 
However, this ``magic angle" is different from MATBG because it does not lead to a flat band in a large portion of the mBZ. 
This is because our $\Gamma$-centered model does not have strong moir\'e effects. 
In the case of cuprates, the quasiparticle dispersion nodes are not near the $\Gamma$- point. 
However, they are still far from the BZ boundaries and thus do not have strong moir\'e effects. 
In spite of this, the QBT still opens possibility of realizing correlated physics, because it leads to a finite density of state at charge neutrality. 
The case of TBMTI is particularly interesting because $\theta_M$ depends on the magnetization. 
Thus for a given TBMTI, one can tune between two separated Dirac nodes to QBT point just by tuning the magnetization using an external magnetic field. 
Thus it opens up possibility of tuning between strongly and weakly correlated phases just by external magnetic field. 
Finite values of $t'$ adds even more richness to this tunability because in that case one can tune between type-II Dirac nodes, a QBT, and type-I Dirac nodes, where in the former two, one can possibly realize strongly correlated electron-hole phases.

\subsection{$M$-point}
If the Dirac dispersion of the isolated MTI is centered around the BZ corner, 
the choice of magnetization along $x$ and $y$ -direction (or the two diagonal planes) are equivalent after an appropriate  $C_{4z}$ rotation. 
Below, we only explicitly consider the case when magnetization is along diagonal plane, i.e. $\vec{m} = (-m/\sqrt{2},m/\sqrt{2},0)$ [See Fig.~\ref{Fig:MTI_model_schematic} (e)]. 
Our main results for magnetization along $x$ or $y$ axis are qualitatively similar with small quantitative corrections. 

For our choice of magnetization the `+' Dirac node  appears at $(\pi-k_D/\sqrt{2},\, \pi-k_D/\sqrt{2})$. 
The clockwise axes rotation to align the Dirac node of the untwisted MTI along $y$-direction, 
modifies the reciprocal lattice vectors as $\vec{g}_1 \rightarrow (\vec{g}_1-\vec{g}_2)/\sqrt{2}$ and $\vec{g}_2 \rightarrow (\vec{g}_1+\vec{g}_2)/\sqrt{2}$. 
Since $\vec{k}_0 = (\vec{g}_1+\vec{g}_2)/2$, the lowest possible value of  $|\vec{G}_{nm}-\vec{k}_0|$ in the rotated basis can be achieved for the set of values $(n,\, m) = (0,\,0), \, (0,\,1),\, (1,\,1),$ and $(-1,\, 1)$. 
We keep these set of tunneling Fourier components in our calculations with equal strength and ignore all the other Fourier components. 

 \begin{figure}[h]
  \includegraphics[width=0.48\textwidth]{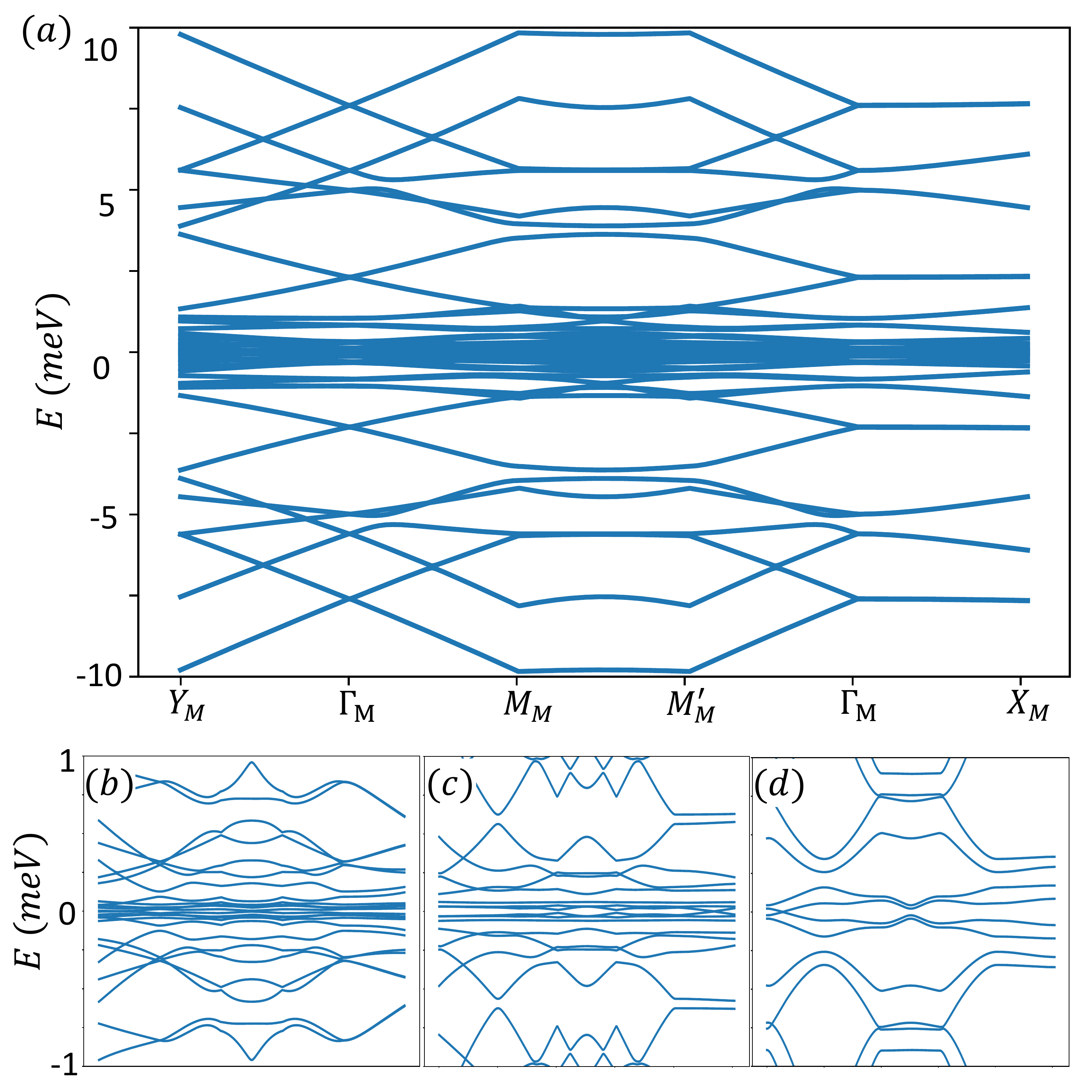}
  \caption{\label{Fig:K_flat}
  (a) Multiple flat bands in small angle ($\theta = 0.4^{\circ}$ here) TBMTI when the monolayer MTI had Dirac dispersion at BZ corner. 
  (b)-(d) Zoom in of the flat bands region as function of twist angles. 
  As twist angles increase, the number of flat bands decrease continuously as depicted here in (b) $\theta = 0.4^{\circ}$, (c) $\theta = 0.6^{\circ}$, and (d) $\theta = 0.8^{\circ}$. 
  The calculations above are done for $v_x = 1$ eV$\cdot\AA$, $v_y = 0.9$ eV$\cdot\AA$, $t_{nm,0} = 0$, $t_{nm,x} = 25 $ meV.
}
\end{figure}

In Fig.~\ref{Fig:K_flat} (a), we show the moir\'e bands of TBMTI for a representative case ($v_x = 1$~eV$\cdot${\AA}, $v_y = 0.9$~eV$\cdot${\AA}, $t_{nm,0} = 0$, $t_{nm,x} = 25 $~meV) at a small twist angle $\theta = 0.4 ^{\circ}$. 
The main observation is the appearance of a large number of extremely flat bands near zero energy. 
As shown in the Fig.~\ref{Fig:K_flat}(b)-(d), we also find that as the twist angle is increased, the number of flat bands continuously decreases as some of the flat bands start to slowly gain dispersion and merge with the higher energy bands. 
Inclusion of finite $t_{nm,0}$ (equivalently $t'$), only adds small quantitative corrections to the low energy bands, which still remain very flat. 
The very high density of flat bands near zero energy and more dispersive and relatively fewer bands at higher energy in Fig.~\ref{Fig:K_flat} also indicate that flat bands are not just a simple artefact of multiple band foldings in the highly reduced mBZ. 
In fact, as we discuss below, all our main observations can be explained due to local zeroth pLL of 2D Dirac fermions due the emergent $SU(2)$ gauge fields.

The appearance of flat bands due to emergent $SU(2)$ vector potential in TBG had been previously studied~\cite{San-Jose2012,Tarnopolsky2019}. Throughout our formulation, we have maintained close resemblance  with TBG, thus the appearance of such flat bands is not a surprise. 
However a few differences exist, as we describe below. 
A more physical understanding can be gained by reinterpreting the $SU(2)$ field as a usual $U(1)$ field. 
For this purpose, we perform a unitary transformation $\psi_s = (\bar{\psi}_t + i\bar{\psi}_b)/\sqrt{2}$ and $\psi_a = (\bar{\psi}_t -i \bar{\psi}_b)/\sqrt{2}$, and then re-scale $\vec{r} \rightarrow \theta \vec{r}$, to obtain
\begin{align}\label{Eq:Ham_TBMTI_U1}
    &H_{TB} = \int d^2\vec{x} \{ \psi^{\dagger}_s[\gamma_x(-i\bar{\partial}_y-A_x) -i\gamma_y \bar{\partial}_x]\psi_s \notag\\
     &+ \psi^{\dagger}_a[\gamma_x(-i\bar{\partial}_y + A_x) -i\gamma_y \bar{\partial}_x]\psi_a 
     + [  \psi^{\dagger}_{s} A_0 \psi_{a} + \text{h.c.} ] \},
\end{align}
where
\begin{subequations}
\begin{align}\label{Eq:U1}
    & A_x(\vec{x}) = - \frac{1}{\theta} \sum_{n,m} t_{nm,x}\sin [(\vec{k}_0+\vec{k}_D-G_{nm})\cdot\hat{z}\times \vec{x}] \\
    & A_0(\vec{x}) = \frac{i}{\theta} \sum_{n,m} t_{nm,x}\cos [(\vec{k}_0+\vec{k}_D-G_{nm})\cdot\hat{z}\times \vec{x}]\gamma_x.
\end{align}
\end{subequations}
Here, without loss of generality, we have assumed that $t_{nm,i}$ are purely real. 
In absence of $A_0$, in this form the system can be interpreted as two 2D Dirac fermions in equal and opposite  periodic $U(1)$ magnetic field. 
This problem has been studied in context of strained graphene~\cite{Guinea2008,Guinea2010,Snyman2009} and topological insulators~\cite{Tang2014} and appearance of pseudo Landau-level (pLL) zero energy states has been shown. 
These pLLs appear due to topological robustness of zeroth Landau level (LL) of Dirac fermions~\cite{Aharonov1979}. 

In the presence of $A_0$, as long as there are multiple tunneling Fourier components involved, it is possible that in certain periodic spatial regions both $A_x$ and $A_0$ vanish simultaneously. Near these regions, up to linear order expansion, $A_x$ takes a form of vector potential of a uniform out of plane magnetic field, while $A_0$ is still vanishingly small (lowest order correction are second order in $A_0$ near these regions).  
Thus locally the system mimics two decoupled Dirac fermions in uniform out-of-plane magnetic field in opposite direction. 
These local zeroth pLL are the origin of the flat bands in the small twist angle limit. Moreover, even including finite $t_{nm,0}$ terms has negligible effect. This is because the two zeroth LLs have opposite spin polarizations and are for this reason not coupled
by tunneling. 

Similar arguments have been previously been used to understand the flat bands in MATBG~\cite{Liu2019}. 
Although qualitatively correct, the above arguments do not present a complete picture of magic angle flat bands in TBG. 
A crucial aspect of magic angles in chiral TBG is that they are a set of isolated angles where the bands near neutrality become exactly flat and isolated from rest of the spectrum. 
In our case, we do not find such isolated magic angles, instead the number of flat bands smoothly decrease as the twist angle is increased.  
As seen in Eq.~\ref{Eq:U1}, the effective $U(1)$ gauge field is inversely proportional to twist angle. 
As a result, with the twist angle increase, the degeneracy of zeroth pLLs continuously decreases. 
Thus we have an evolution of flat bands as shown in Fig.~\ref{Fig:K_flat} (b)-(d), where they smoothly merge into higher energy bands. Such a continuous evolution was also pointed out recently in twisted bilayers of staggered flux square lattice model~\cite{Luo2021}.
It should be noted that even in the TBG with exact chiral symmetry, apart from the isolated instances of magic angles, where exactly two bands (per spin and valley) become perfectly flat, in general there a continuous evolution, where increasing number of nearby bands become very flat as the twist angle is decreased~\cite{Tarnopolsky2019}.

\subsection{$X$ and $Y$-points}
If the Dirac dispersion of the isolated MTI is centered around $X$ or $Y$- point, the system has $M_x$ and  $M_y$ mirror symmetry in absence of magnetization.  
The magnetization along $x (y)$-direction breaks $M_y (M_x)$ mirror symmetry. 
The case of $\vec{k}_0 = X$ with magnetization along $x (y)$ is identical to the case of $\vec{k}_0 = Y$ with magnetization along $y (x)$ after applying $C_{4z}$ rotation. 
In the discussion below, we only consider the magnetization along $x$-direction to stay consistent with the $M_x$ symmetry. 

\begin{figure}[h]
  \includegraphics[width=0.48\textwidth]{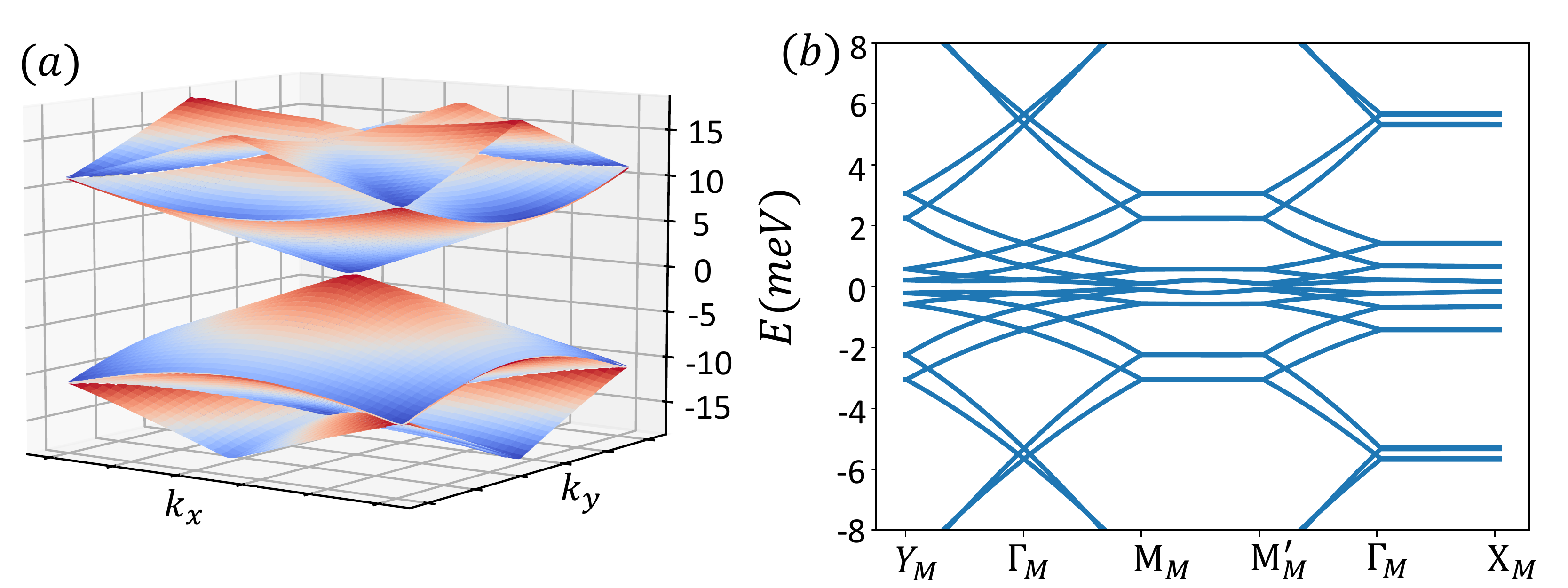}
  \caption{\label{Fig:XY}
  Quadratic band touching and flat bands in TBMTI: (a) Dirac dispersion of MTI centered around $X (Y)$-point with magnetization along $x (y)$ direction. At isolated twist angle the Dirac nodes from top and bottom layers merge to form QBT. (b) Dirac dispersion of MTI centered around $X (Y)$-point with magnetization along $y (x)$ direction. At small twist angles, very flat moir\'e bands appear. 
  The calculations above are done for $v_x = 1$ eV$\cdot\AA$, $v_y = 0.9$ eV$\cdot\AA$, $t_{nm,0} = 0$, $t_{nm,x} = 25 $ meV. The twist angles are $\theta = 0.7^\circ$ and $\theta = 0.6^\circ$ respectively for (a) and (b).
}
\end{figure}

When the Dirac dispersion is centered around $X$-point, the Dirac nodes split along the $y$- direction such that `$+$' node in and in the isolated layer is at $\vec{k} = [\pi,\, -k_D]$. 
In this case, $|\vec{G}_{1,0}-\vec{k}_0| = |\vec{G}_{0,0}-\vec{k}_0|$, the $(n,m) = (0,0),\, (1,0)$, are the two lowest Fourier components that contribute predominantly to tunneling. 
Moreover $(-1,\, 0)$ is related to $(1,\,0)$ component under $M_x\eta_x$ symmetry. 
Thus, we include these three tunneling Fourier components on an equal footing.
In this case, even after including these moir\'e effects, we do not find flat bands. 
However as shown in the Fig.~\ref{Fig:XY} (a), similar to the case of the $\Gamma$-centered model, we find a QBT as functions of twist angle (or magnetization) when the two Dirac nodes merge.

When the Dirac dispersion is centered around $Y$-point, the Dirac nodes split along the $x$-direction such that `$+$' node in the isolated layer is at $\vec{k} = (0,\, - \pi + k_D)$. 
In this case, $|\vec{G}_{0,1}-\vec{k}_0| - |\vec{G}_{0,0}-\vec{k}_0| = 2k_D \ll g_i$. Thus, we keep both these Fourier components at the lowest order in tunneling by assuming equal amplitudes associated with them. 
However, in stark contrast to the $X$- point model considered above, we see appearance of flat bands at small twist angle as shown in the Fig.~\ref{Fig:XY} (b). 
We also find that similar to the $M$ point case, the number of flat bands increase as twist angle is decreased. 
Notice that the case of $Y$-point with magnetization in $y$-direction is identical to the case of $X$-point with magnetization in $x$ discussed in the previous paragraph. 
Thus by simply tuning the magnetization direction from $x$ to $y$, one can tune between flat bands to QBT or separated Dirac nodes. 

The stark contrast between the two cases can be understood by considering the structure of the $SU(2)$ vector potential $\bar{\vec{T}}(\vec{x}) = [0,\,  \bar{T}_x(\vec{x}) ]$. For the $Y$ centered model, the vector potential is simplified to
\begin{align}\label{Eq:vec_pot}
    \bar{\vec{T}}_x(\vec{x}) = 2t_{00,x}\text{e}^{i\theta k_D x} \cos (\theta\pi x) \eta_x \hat{y} , 
\end{align}
which results into a finite spatially varying gauge field similar to $M$ point case, although a little different in exact details. 
This gauge field results into local pLL like states, which is the origin of flat bands as explained earlier. 

In contrast, for the $X$ centered model, the vector potential simplifies to
\begin{align}\label{Eq:vec_pot2}
    \bar{\vec{T}}_x(\vec{x}) = t_{00,x}\text{e}^{-i\theta k_D x} \text{e}^{-i\theta \pi y} [1+2 \cos (2\theta\pi y) ] \eta_x \hat{y} .
\end{align} 

The exponential term above comes from the shift in the Dirac nodes, and its spatial variation happens at a much larger length scale than the moir\'e length-scale. 
Ignoring this slow spatial variation in the exponential term, the associated effective ``magnetic field" $\vec{B} = \nabla \times \bar{\vec{T}}(\vec{x}) \sim 0 $. 
Thus, we do not obtain flat bands in this case because of the absence (very small) effective field. 
Thus for the $X$-centered model, a gauge transformation can remove the vector potential (up to the Dirac node shift term), such that the Hamiltonian formally resembles the case of $\Gamma$- centered model already discussed. 
Thus appearance of a similar QBT is not a surprise. 
Notice that including higher Fourier components can indeed lead to a finite vector potential, and thus flat moir\'e bands for the $X$-centered and $\Gamma$-centered models, likewise. 
The amplitudes, associated with them, are neglected here. 
However, for very small twist angles the higher order Fourier components may indeed become relevant and in principle can lead to moir\'e flat bands for the $\Gamma$ and $X$- centered case as well.

\section{Discussion and Conclusion}

We have shown that highly tunable flat bands can be achieved in TBMTI with a large in-plane magnetization. 
Hence, similar to the MATBG, this system can also be a fertile ground for strongly correlated topological phases.
Unlike graphene, where the underlying monolayer has doubly-degenerate 2D Dirac dispersion because of the spin degeneracy, this case is built upon underlying 
nondegenerate 2D Dirac nodes. 
Thus, potentially, these systems will host correlated phases, that will be distinct from the correlated phases of MATBG.
We would like to stress here that the flat bands, that arise in TBMTI at small twist angles, all originate from the zeroth LL of massless 2D Dirac fermions. 
This implies that, similar to MATBG, we expect strong correlation physics to be manifest at small twist angles, where these flat bands are most prominent, even though TBMTI lacks sharply-defined ``magic angles" of MATBG. 

We conclude by discussing some possible experimental implications of our work. Even though twist angle provides an important tuning knob for strongly correlated phases, 
it has significant inherent limitations, since the twist angle is not easily tunable once the sample is prepared. 
In contrast, in our system the effective interlayer tunneling in the Dirac node projected model depends on the magnetization, which means that the flat bands can be controlled by simply tuning the magnetization. 

Our work shows that the exact details of the moir\'e bands of TBMTI also depends on the position of the original Dirac dispersion of the isolated MTI. 
Current experimentally well established examples of MTIs are chromium-doped (Bi,Sb)$_2$Te$_3$~\cite{Yu2010,Chang2013} and intrinsic MnBi$_2$Te$_4$~\cite{Otrokov_2017,Deng2020}, which have Dirac dispersion centered around the $\Gamma$-point. 
For the $\Gamma$-centered case, the moir\'e effects are negligible and we do not obtain flat bands. 
However, we show that one can still tune between type-I and type-II Dirac nodes and QBT dispersion by tuning twist angle and magnetization. 
Thus it should be possible to achieve these phenomena and the corresponding correlated phases in these materials. 

The more interesting physics related to flat moir\'e bands, occurs when the Dirac dispersions are centered at BZ boundaries ($M$, $X$, and $Y$) points. 
We are not aware of any current MTI with Dirac dispersion centered at BZ boundaries. 
However, this is not a fundamental limitation and possibly with new materials discovery, this avenue will open in experiments. 

Another possible challenge is related to the magnetization orientation. 
The magnetization in chromium-doped (Bi,Sb)$_2$Te$_3$ and MnBi$_2$Te$_4$ tends to have easy axis out of plane anisotropy. 
However, a small external field can easily overcome this and point the magnetization in-plane. 
In spite of this, a finite out-of-plane magnetization is likely to exist in experiments. 
This out of plane magnetization also breaks the mirror symmetry that protects the Dirac nodes. 
As a result, the effect of mirror symmetry breaking perturbations can qualitatively be modeled by a small out of plane magnetization. 
Such perturbations typically open up a gap that are expected to have a finite Chern number. 
However, we still expect the flat bands to exist (although gapped). 
Thus the system can potentially also realize flat Chern bands that can be a breeding ground for fractional Chern insulators~\cite{Lian2020}. 

Recently, it was shown that variation of the interlayer distance in twisted bilayers of topological insulators can lead to local band inversion at moir\'e length scale, thus leading to topologically distinct local phases~\cite{Tateishi2021}. In our work, we have excluded such local variations, but expect them to add further richness to the possible correlated phase in this system, which should be an interesting future direction.

Finally, similar to the isolated valleys of monolayer graphene, we have assumed the two Dirac nodes within a layer to be isolated.  
However, neither the momentum separation nor the energy barriers in this case are as large as in graphene. 
Thus possibly the intervalley terms are important. 
In real systems, we expect a sweet spot of magnetization, tunnelings, and Fermi velocities, where our  approximation is valid. 
The inter-valley effects and detailed phase diagram with magnetization and tunnelings are subject for future studies.

\section{Acknowledgement}
This work was supported by Center for Advancement of Topological Semimetals, an Energy Frontier Research Center funded by the U.S. Department of Energy Office of Science, Office of Basic Energy Sciences, through the Ames Laboratory under
contract DE-AC02-07CH11358. 
Research at Perimeter Institute is supported in part by the Government of Canada through the Department of Innovation, Science and Economic Development and by the Province of Ontario through the Ministry
of Economic Development, Job Creation and Trade. 
We gratefully acknowledge the computing resources provided on Bebop and Blues,  high-performance computing clusters operated by the Laboratory Computing Resource Center at Argonne National Laboratory.

\appendix

\section{Derivation of the bilayer model\label{Sec:App_bilayer_derivation}}
Here we discuss the bilayer MTI model and the steps to obtain the low energy effective Hamiltonian in Eq.~\ref{Eq:Ham_MTI_transformed} and Eq.~\ref{Eq:Ham_BL_approx}. 
We start with a given lattice model Hamiltonian with Dirac dispersion near a TRIM point $\vec{k}_0$ and a specific in-plane magnetization direction perpendicular to a mirror plane. 
Next, we expand the Hamiltonian to linear order around $\vec{k}_0$. 
In Table~\ref{Table:kdotp}, we list possible combinations of a TRIM point,  magnetization, and relevant linearly expanded model. 
All these cases can generically be described by the simple Hamiltonian: 
\begin{align}\label{Eq:App_MTI_linear}
    h(\vec{k}) = v( k_y\sigma_x - k_x\sigma_y)\tau_z + m \sigma_x +\Delta\tau_x .
\end{align}
Notice that the model near $Y$-point with magnetization in $y (x)$-direction and model near $X$-point with magnetization in $x (y)$-direction are equivalent by space rotation. 
Similarly, the model near $\Gamma$ or $M$-point with magnetization along a diagonal plane can be written in the above form by performing a space rotation and Pauli matrix rotation by $\pi/4$. 
The space rotation also leads to rotation of the reciprocal lattice vector $\vec{g}_1$ and $\vec{g}_2$ of the square lattice. 
This fact will be important when considering the twisted bilayer configuration, since the interlayer tunneling follows the translation symmetry of the underlying square lattice model. 
We also mention that in the form of Eq.~\ref{Eq:App_MTI_linear} for all the considered case, the mirror symmetry is always under $yz$-mirror plane.

The derivation after this follows by maintaining a close analogy to Dirac dispersion in graphene. 
With this in mind, we transform the Hamiltonian in Eq.~\ref{Eq:App_MTI_linear} to the form
\begin{widetext}
\begin{align}\label{Eq:App_Ham_MTI_transformed}
    \bar{h}(\vec{k}) = \begin{pmatrix}
    (m+\sqrt{v^2k^2_y+\Delta^2})\gamma_x-vk_x\gamma_y & 0_{2\times 2} \\
    0_{2\times 2} & (m-\sqrt{v^2k^2_y+\Delta^2})\gamma_x+vk_x\gamma_y
    \end{pmatrix}
\end{align}
using the similarity transformation $\bar{h} = S^{-1} h S$ ,
where
\begin{align}\label{Eq:App_Transformataion}
    S = \begin{pmatrix}
    0 & \frac{vk_y+\sqrt{v^2k^2_y + \Delta^2}}{\Delta} & 0 & \frac{vk_y-\sqrt{v^2k^2_y + \Delta^2}}{\Delta} \\
    \frac{(m-ivk_x+\sqrt{v^2k^2_y+\Delta^2})(\Delta^2+vk_y(vk_y+\sqrt{v^2k^2_y+\Delta^2}))}{\Delta\sqrt{v^2k^2_y+\Delta^2}(m+ivk_x+\sqrt{v^2k^2_y+\Delta^2})} & 0 & \frac{vk_y-\sqrt{v^2k^2_y + \Delta^2}}{\Delta} & 0 \\
    \frac{m-ivk_x+\sqrt{v^2k^2_y+\Delta^2}}{m+ivk_x\sqrt{v^2k^2_y+\Delta^2}} & 0 & 1 & 0\\
    0 & 1 & 0 & 1
    \end{pmatrix}
\end{align}
\end{widetext}

The transformed monolayer MTI Hamiltonian in Eq.~\ref{Eq:App_Ham_MTI_transformed} has decoupled high energy and low energy sector in the upper and lower diagonal blocks. 

\begin{widetext}
\begin{table*}\label{Table:kdotp}
\caption{Effective model near Dirac point of isolated MTI layer }
\begin{center}\label{Table:2}
 \begin{tabular}{|c | c | c | c | c |} 
 \hline
 TRIM  & $\vec{m}$ & $H(\vec{k})$ & Mirror symmetry  \\ [0.5ex]
 \hline
  \multirow{2}{*} {$\Gamma$} &  $(m,\,0,\,0)$   &  $v(k_y\sigma_x - k_x\sigma_y)\tau_z + m\sigma_x + \Delta \tau_x $  & $M_x: \sigma_y H(k_x,-k_y)\sigma_y = H(k_x,k_y)$  \\ 
  & $\frac{1}{\sqrt{2}} (m,\,m,\,0)$ &  $v(k_y\sigma_x - k_x\sigma_y)\tau_z + \frac{m}{\sqrt{2}}(\sigma_x+\sigma_y) + \Delta \tau_x$  & $M_d: \frac{\sigma_x+\sigma_y}{\sqrt{2}} H(-k_y,-k_x)\frac{\sigma_x+\sigma_y}{\sqrt{2}} = H(k_x,k_y)$ \\
 \hline
  \multirow{2}{*} {$X$} & $(m,\,0,\,0)$ &  $v(k_y\sigma_x + k_x\sigma_y)\tau_z + m\sigma_x + \Delta \tau_x $ & $M_x: \sigma_y H(k_x,-k_y)\sigma_y = H(k_x,k_y)$  \\
  & $(0,\,m,\,0)$ &  $v(k_y\sigma_x + k_x\sigma_y)\tau_z + m\sigma_y + \Delta \tau_x $ & $M_y: \sigma_x H(-k_x,k_y)\sigma_x = H(k_x,k_y)$  \\
 \hline
  \multirow{2}{*} {$Y$} & $(m,\,0,\,0)$ & $v(-k_y\sigma_x - k_x\sigma_y)\tau_z + m\sigma_x + \Delta \tau_x $ & $M_x: \sigma_y H(k_x,-k_y)\sigma_y = H(k_x,k_y)$    \\
  & $(0,\,m,\,0)$ & $v(-k_y\sigma_x - k_x\sigma_y)\tau_z + m\sigma_y + \Delta \tau_x $ & $M_y: \sigma_x H(-k_x,k_y)\sigma_x = H(k_x,k_y)$    \\
 \hline
 \multirow{3}{*} {$M$} & $(m,\,0,\,0)$ & $v(-k_y\sigma_x + k_x\sigma_y)\tau_z + m\sigma_x + \Delta \tau_x $ & $M_x: \sigma_y H(k_x,-k_y)\sigma_y = H(k_x,k_y)$    \\
  & $(0,\,m,\,0)$ & $v(-k_y\sigma_x + k_x\sigma_y)\tau_z + m\sigma_y + \Delta \tau_x $ & $M_y: \sigma_x H(-k_x,k_y)\sigma_x = H(k_x,k_y)$    \\
  & $\frac{1}{\sqrt{2}}(m,\,m,\,0)$ & $v(-k_y\sigma_x+k_x\sigma_y)\tau_z + \frac{m}{\sqrt{2}}(\sigma_x+\sigma_y) + \Delta \tau_x$ & $M_d: \frac{\sigma_x+\sigma_y}{\sqrt{2}} H(-k_y,-k_x)\frac{\sigma_x+\sigma_y}{\sqrt{2}} = H(k_x,k_y)$  \\
 \hline
\end{tabular}
\end{center}
\end{table*}
\end{widetext}

Now we consider bilayers of MTI following Eq.~\ref{Eq:Ham_bilayer} and~\ref{Eq:Ham_tunneling} and perform the above mentioned similarity transformation, which transforms the tunneling Hamiltonian as
\begin{widetext}
\begin{align}\label{Eq:App_tunnel_transformed}
    \bar{T} = \begin{pmatrix}
    t' & \frac{t\Delta(m+\sqrt{v^2k^2_y+\Delta^2}+iv k_x)}{2\sqrt{v^2k^2_y+\Delta^2}(m+\sqrt{v^2k^2_y+\Delta^2}-ivk_x)} & 0 & \frac{t\Delta(m+\sqrt{v^2k^2_y+\Delta^2}+iv k_x)}{2\sqrt{v^2k^2_y+\Delta^2}(m+\sqrt{v^2k^2_y+\Delta^2}-ivk_x)} \\
    \frac{t\Delta(m+\sqrt{v^2k^2_y+\Delta^2}-iv k_x)}{2\sqrt{v^2k^2_y+\Delta^2}(m+\sqrt{v^2k^2_y+\Delta^2}+ivk_x)} & t' & \frac{t\Delta}{2\sqrt{v^2k^2_y+\Delta^2}} & 0 \\
    0 & -\frac{t\Delta}{2\sqrt{v^2k^2_y+\Delta^2}} & t' & -\frac{t\Delta}{2\sqrt{v^2k^2_y+\Delta^2}} \\
    -\frac{t\Delta(m+\sqrt{v^2k^2_y+\Delta^2}-iv k_x)}{2\sqrt{v^2k^2_y+\Delta^2}(m+\sqrt{v^2k^2_y+\Delta^2}+ivk_x)} & 0 & -\frac{t\Delta}{2\sqrt{v^2k^2_y+\Delta^2}} & t'
    \end{pmatrix}.
\end{align}
\end{widetext}
Here the upper (lower) $2\times 2$ diagonal entries represent interlayer tunneling between the high (low) energy sectors of top and bottom MTI layers. 
The off-diagonal $2\times 2$ block represent tunneling between the low energy sector of one layer to the high energy sector of the other layer. 
Because the high energy sector is separated by $2m$ in energy from low energy sector, this inter-sector tunneling only makes very small quantitative correction to the low energy sector. More importantly it does not break any symmetry of the low energy sector and the Dirac nodes remain protected. 
Thus, from now on we ignore the off-diagonal $2\times 2$ tunneling block in Eq.~\ref{Eq:App_tunnel_transformed}, and as a result we only consider the low energy sector in isolation, given by
\begin{widetext}
\begin{align}\label{Eq:App_Ham_BL_low}
    H_{BL,L}(\vec{k}) =  
    \begin{pmatrix}
    (m-\sqrt{v^2k^2_y+\Delta^2})\gamma_x+vk_x \gamma_y & t'\eta_0 -\frac{t\Delta}{2\sqrt{v^2k^2_y+\Delta^2}} \eta_x \\
    t'\eta_0 -\frac{t\Delta}{2\sqrt{v^2k^2_y+\Delta^2}} \eta_x & (m-\sqrt{v^2k^2_y+\Delta^2})\gamma_x+vk_x \gamma_y
    \end{pmatrix}.
\end{align}
\end{widetext}
As mentioned in the main text, here $\eta_i$ and $\gamma_i$ are Pauli matrices to give $2\times 2$ matrix structure. 

When measured from $\vec{k}_0$, the monolayer has Dirac nodes at $\pm \vec{k}_D = (0,\pm  \sqrt{m^2-\Delta^2}/v)$. Assuming, the two Dirac nodes to be decoupled, we expand to the linear order in momentum near one of the Dirac node location [ to be specific, we choose $\vec{k}_D = (0, \sqrt{m^2-\Delta^2}/v)$ ]  to obtain
\begin{widetext}
\begin{align}\label{Eq:App_Ham_BL_low_approx}
    H_{+,BL}(\vec{k}) = \begin{pmatrix}
    \biggl (1-\frac{\Delta^2}{2m^2} \biggr )vk_y\gamma_x + vk_x \gamma_y & t'\eta_0 -\frac{t\Delta}{2 m} \eta_x \\
    t'\eta_0 -\frac{t\Delta}{2 m} \eta_x & \biggl (1-\frac{\Delta^2}{2m^2} \biggr )vk_y\gamma_x + vk_x \gamma_y
    \end{pmatrix}, 
\end{align}
\end{widetext}
which is $\chi = +$ of Eq.~\ref{Eq:Ham_BL_approx} in the main text.

\section{Derivation of twisted bilayer model\label{Sec:App_TB_derivation}}
In this section, we derive the continuum model for TBMTI near the Dirac node. We closely follow the formulation in Ref.~\cite{Balents2019}. 
Here we highlight some important steps first. 
It is more intuitive to work in the real space. 
Assuming two decoupled Dirac nodes from a single MTI layer, general real space wavefunction can be described as
\begin{align}\label{Eq:App_real_space_coeff}
    c(\vec{r}) = \psi_+(\vec{r}) \text{e}^{i(\vec{k}_0+\vec{k}_D)\cdot\vec{r}} + \psi_-(\vec{r}) \text{e}^{i(\vec{k}_0-\vec{k}_D)\cdot\vec{r}} ,
\end{align}
where $\psi_{\pm}$ are the two component electron annihilation operators at two valleys and $\vec{r}$ is a position coordinate in an undeformed solid (in this case, before twist). 
The real space Hamiltonian near valley $+$ (near $\vec{k}_0+\vec{k}_D$ Dirac node) in this approximation can be written as
\begin{align}\label{Eq:App_Weyl_Ham}
    H_D = -i\sum_{\mu=1,2}\int d^2 \vec{r} \psi^{\dagger} v_\mu\gamma_\mu  \frac{\partial}{\partial\vec{r}_\mu}\psi, 
\end{align} 
which is simply the real space representation of the diagonal $2\times 2$ block of Hamiltonian in Eq.~\ref{Eq:App_Ham_BL_low_approx} and written as $\vec{k}_0+\vec{k}_D$ as the momentum space origin. 
We have also suppressed the valley subscript `$+$'.

We can account for small twist in a more general form by considering a small real space deformation field $\vec{u}(\vec{x})$ that results in transformation
\begin{align}\label{Eq:App_deform}
    \vec{x} = \vec{r} + \vec{u}(\vec{x}),
\end{align} 
such that the location $\vec{r}$ in the solid is moved to $x$ in the laboratory frame after deformation.
Under this deformation
\begin{align}\label{Eq:App_deform_wave1}
    c(\vec{x}) = \sqrt{\biggl |\text{det} \frac{\partial r_\mu}{\partial x_\nu} \biggr | } c(\vec{r}(\vec{x})) \sim \sqrt{1-\nabla\cdot \vec{u} }\, c(\vec{r}(\vec{x})),
\end{align}
we obtain
\begin{align}\label{Eq:App_deform_wave2}
    c(\vec{x}) = \psi(\vec{x}) \text{e}^{i(\vec{k}_0+\vec{k}_D)\cdot\vec{x}}.
\end{align}
This gives relation between the electron annihilation operators
\begin{align}\label{Eq:App_deform_annhilation}
    \psi(\vec{r}) = \frac{\psi(\vec{x})}{\sqrt{1-\nabla\cdot u}}\text{e}^{i(\vec{k}_0+\vec{k}_D)\cdot\vec{u}(\vec{x})}
\end{align}
and the integration measure is modified as 
\begin{align}\label{Eq:App_deform_int}
    d^2\vec{r} \sim d^2 \vec{x} (1-\nabla\cdot\vec{u})
\end{align}

Substituting Eq.~\ref{Eq:App_deform_annhilation} and~\ref{Eq:App_deform_int} in Eq.~\ref{Eq:App_Weyl_Ham} and taking the small deformation approximation by ignoring the $O(\partial^2 u)$ and $O((\partial u)^2)$ terms, we obtain
\begin{align}\label{Eq:App_deform_Weyl_Ham}
    & H_D = \sum_{\mu=1,2} \int d^2 \vec{x} \psi^\dagger \biggl [-i \biggl ( v_{\mu}\gamma_{\mu} + \frac{\partial u_{\mu}}{\partial x_{\nu} } v_{\nu}\gamma_\nu \biggr ) \frac{\partial}{\partial x_{\mu}} \notag\\
    & \hspace{3cm}+  (\vec{k}_0+\vec{k}_D) \cdot \frac{\partial \vec{u}} {\partial x_{\mu}}  v_{\mu} \gamma_\mu \biggr ] \psi
\end{align}
as the Hamiltonian near the $\chi=+$ Dirac node under the small deformation field. 

Twisted bilayers of MTI are then constructed by considering two copies of deformed Dirac Hamiltonian of Eq.~\ref{Eq:App_deform_Weyl_Ham}, which are twisted at rigid angles $\pm \theta/2$, such as the deformation field of rigid twist is given by
\begin{align}\label{Eq:App_twist_deform}
    \vec{u}_t = -\vec{u}_b = \frac{\vec{u}}{2} = \frac{\theta}{2}\hat{z}\times \vec{x}.
\end{align}
Such deformation field associated with rigid twist has has no divergence, i.e. $\vec{\nabla}\cdot\vec{u} = 0$, and
\begin{align}\label{Eq:App_deform_Weyl_shift}
    \frac{\partial \vec{u}}{\partial x_1 } =  ( 0,\, \theta  ),\quad 
    \frac{\partial \vec{u}}{\partial x_2} = (-\theta,\, 0 ),
\end{align}
which results in the shift of Dirac nodes in top and bottom layer given by 
\begin{align}\label{Eq:App_deform_Weyl_shift2}
    d\vec{k}_{D,t} = - d\vec{k}_{D,b} = \frac{\theta}{2}  (k_{0y}+k_{Dy},\, -k_{0x}-k_{Dx}).
\end{align}

Finally, relabelling $(\partial/\partial x_1,\, \partial/\partial x_2) \equiv (\partial_x, \, \partial_y)$, we obtain the twisted bilayer Hamiltonian in the Eq.~\ref{Eq:Ham_TBMTI} of the main text. 
For the implementation of the $M_x\eta_x$ symmetry, it is convenient to remove the Dirac node shift terms in the individual layer by gauge transformation $\bar{\psi}_{t/b} = \text{e}^{\pm i r \theta \gamma_z/4} \text{e}^{\mp i\theta \hat{z}\times(\vec{k}_0+\vec{k}_D)\cdot \vec{x}/2}\psi_{t/b}$ and write the Hamiltonian as, 
\begin{align}\label{Eq:App_Ham_TBMTI_transform}
    &H_{TB} = \int d^2\vec{x} [ \bar{\psi}^{\dagger}_t(-i\bm{\gamma}\cdot \bar{\bm{\nabla}} ) \bar{\psi}_t + \bar{\psi}^{\dagger}_b (-i\bm{\gamma}\cdot \bar{\bm{\nabla}}  ) \bar{\psi}_b \notag\\
     &+ (  \text{e}^{-i\theta \hat{z}\times (\vec{k}_0+\vec{k}_D)\cdot\vec{x}}\bar{\psi}^{\dagger}_{t} \text{e}^{i\frac{r\theta}{4}\gamma_z} T(\vec{x}) \text{e}^{i\frac{r\theta}{4}\gamma_z}\bar{\psi}_{b} + \text{h.c.} ) ].
\end{align} 

Under the $M_x\eta_x$ operation $\psi_t(x,y)\rightarrow \psi_b(-x,y)$, thus for the above Hamiltonian be  symmetric, the tunneling Hamiltonian must follow
\begin{align}\label{Eq:App_mirror_tunnel}
    T(x,y) = T^{\dagger}(-x,y), 
\end{align}
which leads to the condition $t_{nm,x/0} = t^{\ast}_{-nm,x/0}$ stated in the main text. 

Now, the final step is to write down the form of interlayer tunneling Hamiltonian $T(\vec{x})$.  
This is done by incorporating the $M_x\eta_x$ symmetry above and using standard approximations of BM model. 
To do so, the interlayer tunneling can be expanded over the reciprocal momenta of the underlying square lattice of the single layer of MTI, since the overall shift of twisted bilayer by the lattice vectors of the underlying square-lattice is invariant. 
The periodic moir\'e potential couples a momentum state from one layer to a different momentum state in another layer. 
The tunneling matrix elements in the momentum space are
\begin{align}\label{Eq:App_tunneling_matrix_k}
    & T_{i,j}(\vec{k}_t,\vec{k}_b) = \sum_{\vec{R}_t,\vec{R}_b} \text{e}^{-i\vec{k}_t\cdot\vec{R}_t} \text{e}^{i\vec{k}_b\cdot\vec{R}_b} T_{i,j}(\vec{R}_t-\vec{R}_b),
\end{align}

where $\vec{k}_t$ and $\vec{k}_b$ are the momentum states and $\vec{R}_t$ and $\vec{R}_b$ take values over the lattice sites in the top and bottom layers respectively. 
Here, we have made the standard moir\'e tight-binding assumption that the crystal locally has the translation invariance.
Using Fourier transform 
\begin{align}\label{Eq:App_tunneling_matrix_kk}
    & T_{i,j}(\vec{k}_t,\vec{k}_b) =\sum_{\vec{R}_t,\vec{R}_b}\sum_{\vec{k}}  \text{e}^{i(\vec{k}-\vec{k}_t)\cdot\vec{R}_t} \text{e}^{-i\vec{R}_b\cdot(\vec{k}-\vec{k}_b)} T_{i,j}(\vec{k}) \notag\\
    & \hspace{1.5cm} = \sum_{\vec{G}_t,\vec{G}_b} \delta_{\vec{k}_t+\vec{G}_t,\vec{k}_b+\vec{G}_b}  T_{i,j}(\vec{k}_t+\vec{G}_t), 
\end{align}
where we decompose the momenta in the above expression in the following manner:
\begin{subequations}\label{Eq:App_momentum_decomp}
\begin{align}
    & \vec{k}_t = d\vec{k}_{D,t} + d\vec{k}_t, \\ 
    & \vec{k}_b = d\vec{k}_{D,b} + d\vec{k}_b, \\
    & \vec{G}_t = \vec{G}_{nm,t} + d\vec{G}_t - \vec{k}_0 - \vec{k}_D,  \\
    & \vec{G}_b = \vec{G}_{n'm',b} + d\vec{G}_b - \vec{k}_0 - \vec{k}_D.  
\end{align}
\end{subequations} 
The new momenta introduced here are as follows: $d\vec{k}_t$ and $d\vec{k}_b$ denote the momenta relative to shifted Dirac nodes in the top and bottom layers respectively; $\vec{G}_{nm,t} = (n\vec{g}_1+m\vec{g}_2)$ and $\vec{G}_{n'm',b} = (n'\vec{g}_1+m'\vec{g}_2)$ for integers $n,\,m,\,n',\,m'$ denote the reciprocal lattice vectors before twist; $d\vec{G}_t$ and $d\vec{G}_b$ denote the change in the reciprocal vector after twisting the respective layers. 
Substituting Eq.~\ref{Eq:App_momentum_decomp} in Eq.~\ref{Eq:App_tunneling_matrix_kk}, we obtain
\begin{align}\label{Eq:App_tunnleing_matrix_kk2}
    &T_{i,j}(\vec{k}_t,\vec{k}_b) \equiv T_{i,j}(d\vec{k}_{D,t}+d\vec{k}_t,d\vec{k}_{D,b}+d\vec{k}_b) \notag\\ &=\sum_{\substack{n,m\\n',m'}}\sum_{\substack{d\vec{G}_t\\d\vec{G}_b}} 
    \delta_{\vec{G}_{nm,t}-\vec{G}_{n'm',b}, d\vec{G}_b-d\vec{G}_t+d\vec{k}_{D,b}-d\vec{k}_{D,t}+d\vec{k}_b-d\vec{k}_t} \notag\\
    &\hspace{0.5cm} T_{i,j}(\vec{G}_{nm,t}-\vec{k}_0 -\vec{k}_D+d\vec{G}_t +d\vec{k}_{D,t} +d\vec{k}_t).
\end{align}
Since for small twist angles $G_{nm,t}$ and $G_{n'm',b}$ wavevectors are much larger than all the other wavevectors (except when $n=m=0$), the $\delta$-function identity is only satisfied when
\begin{align}\label{Eq:App_delta_func_simp}
    & \delta_{\vec{G}_{nm,t}-\vec{G}_{n'm',b}, d\vec{G}_b-d\vec{G}_t+d\vec{k}_{D,b}-d\vec{k}_{D,t}+d\vec{k}_b-d\vec{k}_t} \notag\\
    &\hspace{1cm} = \delta_{\vec{0}, d\vec{G}_b-d\vec{G}_t+d\vec{k}_{D,b}-d\vec{k}_{D,t}+d\vec{k}_b-d\vec{k}_t}.
\end{align}
Using $d\vec{k}_D = d\vec{k}_{D,t}-d\vec{k}_{D,b}$, $d\vec{G}_t-d\vec{G}_b = d\vec{G} = \theta (G_{nm,y},\,-G_{nm,x})$, and substituting Eq.~\ref{Eq:App_delta_func_simp} in Eq.~\ref{Eq:App_tunneling_matrix_kk3}, we obtain
\begin{align}\label{Eq:App_tunneling_matrix_kk3}
    & T_{i,j}(d\vec{k}_{D,t}+d\vec{k}_t,d\vec{k}_{D,b}+d\vec{k}_b) =\sum_{n,m} 
    \delta_{d\vec{k}_t+d\vec{G}+d\vec{k}_D ,d\vec{k}_b} \notag\\
    & T_{i,j}(\vec{G}_{nm,t}-\vec{k}_0 - \vec{k}_D + d\vec{k}_t +\frac{1}{2}(d\vec{G} +d\vec{k}_D) ) \notag\\
    & \implies T_{i,j}(\vec{k}_t,\vec{k}_b) =\sum_{n,m} 
    \delta_{\vec{k}_t+d\vec{G} ,\vec{k}_b} \notag\\
    & T_{i,j}(\vec{G}_{nm,t}-\vec{k}_0 - \vec{k}_D + d\vec{k}_t +\frac{1}{2}(d\vec{G} +d\vec{k}_D) ).
\end{align}

Since the interlayer spacing is much larger than the inter-atomic spacing, tunneling decays slowly in position space and thus is strongly peaked in the momentum space in first few Brillouine zones. 
This is a standard approximation of the BM model~\cite{Bistritzer2011}.
Using the same approximation, $T_{i,j}(\vec{Q})$ to be finite only for some small momenta $\vec{Q}$. 
Hence, we only consider the smallest possible values of $\vec{G}_{nm,t} - \vec{k}_0$. 
Finally, based on this approximation and Eq.~\ref{Eq:App_tunneling_matrix_kk3}, we arrive at the tunneling expression
\begin{align}\label{Eq:App_tunneling_expression_twist}
    & T(\vec{x}) = \sum_{n,m} \text{e}^{-i\theta (G_{nm,y}x-G_{nm,x}y)} T_{nm} \notag\\
    &\hspace{0.75cm} =  \sum_{n,m} \text{e}^{-i\theta \vec{G}_{n,m}\cdot (\hat{z}\times\vec{x})} T_{nm} .
\end{align}

In the limit of zero twist angel, for the above tunneling expression to resemble Eq.~\ref{Eq:Ham_tunnel_transformed} in main text, we choose 
\begin{align}\label{Eq:App_tunnel_choose}
    T_{nm} = t_{nm,0}\eta_0 + t_{nm,x}\eta_x.
\end{align}

Finally, using Eqs.~\ref{Eq:App_tunneling_expression_twist},~\ref{Eq:App_tunnel_choose}, the interlayer tunneling takes the form
\begin{align}\label{Eq:App_tunnel_final}
    T(\vec{x}) = \sum_{n, m\in \mathbb{Z}}  \text{e}^{-i\vec{G}_{nm}\cdot\vec{u}} (t_{nm,0}\eta_0+t_{nm,x}\eta_x), 
\end{align}
with the condition that $t_{nm,i} = t^{\ast}_{-nm,i}$, which comes from the mirror symmetry condition specified in Eq.~\ref{Eq:App_mirror_tunnel}.

\bibliography{bibliography}

\begin{thebibliography}{49}%
\makeatletter
\providecommand \@ifxundefined [1]{%
 \@ifx{#1\undefined}
}%
\providecommand \@ifnum [1]{%
 \ifnum #1\expandafter \@firstoftwo
 \else \expandafter \@secondoftwo
 \fi
}%
\providecommand \@ifx [1]{%
 \ifx #1\expandafter \@firstoftwo
 \else \expandafter \@secondoftwo
 \fi
}%
\providecommand \natexlab [1]{#1}%
\providecommand \enquote  [1]{``#1''}%
\providecommand \bibnamefont  [1]{#1}%
\providecommand \bibfnamefont [1]{#1}%
\providecommand \citenamefont [1]{#1}%
\providecommand \href@noop [0]{\@secondoftwo}%
\providecommand \href [0]{\begingroup \@sanitize@url \@href}%
\providecommand \@href[1]{\@@startlink{#1}\@@href}%
\providecommand \@@href[1]{\endgroup#1\@@endlink}%
\providecommand \@sanitize@url [0]{\catcode `\\12\catcode `\$12\catcode
  `\&12\catcode `\#12\catcode `\^12\catcode `\_12\catcode `\%12\relax}%
\providecommand \@@startlink[1]{}%
\providecommand \@@endlink[0]{}%
\providecommand \url  [0]{\begingroup\@sanitize@url \@url }%
\providecommand \@url [1]{\endgroup\@href {#1}{\urlprefix }}%
\providecommand \urlprefix  [0]{URL }%
\providecommand \Eprint [0]{\href }%
\providecommand \doibase [0]{http://dx.doi.org/}%
\providecommand \selectlanguage [0]{\@gobble}%
\providecommand \bibinfo  [0]{\@secondoftwo}%
\providecommand \bibfield  [0]{\@secondoftwo}%
\providecommand \translation [1]{[#1]}%
\providecommand \BibitemOpen [0]{}%
\providecommand \bibitemStop [0]{}%
\providecommand \bibitemNoStop [0]{.\EOS\space}%
\providecommand \EOS [0]{\spacefactor3000\relax}%
\providecommand \BibitemShut  [1]{\csname bibitem#1\endcsname}%
\let\auto@bib@innerbib\@empty
\bibitem [{\citenamefont {Cao}\ \emph {et~al.}(2018{\natexlab{a}})\citenamefont
  {Cao}, \citenamefont {Fatemi}, \citenamefont {Demir}, \citenamefont {Fang},
  \citenamefont {Tomarken}, \citenamefont {Luo}, \citenamefont
  {Sanchez-Yamagishi}, \citenamefont {Watanabe}, \citenamefont {Taniguchi},
  \citenamefont {Kaxiras}, \citenamefont {Ashoori},\ and\ \citenamefont
  {Jarillo-Herrero}}]{Cao2018a}%
  \BibitemOpen
  \bibfield  {author} {\bibinfo {author} {\bibfnamefont {Y.}~\bibnamefont
  {Cao}}, \bibinfo {author} {\bibfnamefont {V.}~\bibnamefont {Fatemi}},
  \bibinfo {author} {\bibfnamefont {A.}~\bibnamefont {Demir}}, \bibinfo
  {author} {\bibfnamefont {S.}~\bibnamefont {Fang}}, \bibinfo {author}
  {\bibfnamefont {S.~L.}\ \bibnamefont {Tomarken}}, \bibinfo {author}
  {\bibfnamefont {J.~Y.}\ \bibnamefont {Luo}}, \bibinfo {author} {\bibfnamefont
  {J.~D.}\ \bibnamefont {Sanchez-Yamagishi}}, \bibinfo {author} {\bibfnamefont
  {K.}~\bibnamefont {Watanabe}}, \bibinfo {author} {\bibfnamefont
  {T.}~\bibnamefont {Taniguchi}}, \bibinfo {author} {\bibfnamefont
  {E.}~\bibnamefont {Kaxiras}}, \bibinfo {author} {\bibfnamefont {R.~C.}\
  \bibnamefont {Ashoori}}, \ and\ \bibinfo {author} {\bibfnamefont
  {P.}~\bibnamefont {Jarillo-Herrero}},\ }\bibfield  {title} {\enquote
  {\bibinfo {title} {Correlated insulator behaviour at half-filling in
  magic-angle graphene superlattices},}\ }\href {\doibase 10.1038/nature26154}
  {\bibfield  {journal} {\bibinfo  {journal} {Nature}\ }\textbf {\bibinfo
  {volume} {556}},\ \bibinfo {pages} {80--84} (\bibinfo {year}
  {2018}{\natexlab{a}})}\BibitemShut {NoStop}%
\bibitem [{\citenamefont {Cao}\ \emph {et~al.}(2018{\natexlab{b}})\citenamefont
  {Cao}, \citenamefont {Fatemi}, \citenamefont {Fang}, \citenamefont
  {Watanabe}, \citenamefont {Taniguchi}, \citenamefont {Kaxiras},\ and\
  \citenamefont {Jarillo-Herrero}}]{Cao2018}%
  \BibitemOpen
  \bibfield  {author} {\bibinfo {author} {\bibfnamefont {Y.}~\bibnamefont
  {Cao}}, \bibinfo {author} {\bibfnamefont {V.}~\bibnamefont {Fatemi}},
  \bibinfo {author} {\bibfnamefont {S.}~\bibnamefont {Fang}}, \bibinfo {author}
  {\bibfnamefont {K.}~\bibnamefont {Watanabe}}, \bibinfo {author}
  {\bibfnamefont {T.}~\bibnamefont {Taniguchi}}, \bibinfo {author}
  {\bibfnamefont {E.}~\bibnamefont {Kaxiras}}, \ and\ \bibinfo {author}
  {\bibfnamefont {P.}~\bibnamefont {Jarillo-Herrero}},\ }\bibfield  {title}
  {\enquote {\bibinfo {title} {Unconventional superconductivity in magic-angle
  graphene superlattices},}\ }\href {\doibase 10.1038/nature26160} {\bibfield
  {journal} {\bibinfo  {journal} {Nature}\ }\textbf {\bibinfo {volume} {556}},\
  \bibinfo {pages} {43--50} (\bibinfo {year} {2018}{\natexlab{b}})}\BibitemShut
  {NoStop}%
\bibitem [{\citenamefont {dos Santos}\ \emph {et~al.}(2007)\citenamefont {dos
  Santos}, \citenamefont {Peres},\ and\ \citenamefont {Neto}}]{Santos2007}%
  \BibitemOpen
  \bibfield  {author} {\bibinfo {author} {\bibfnamefont {J.~M. B.~Lopes}\
  \bibnamefont {dos Santos}}, \bibinfo {author} {\bibfnamefont {N.~M.~R.}\
  \bibnamefont {Peres}}, \ and\ \bibinfo {author} {\bibfnamefont
  {A.~H.~Castro}\ \bibnamefont {Neto}},\ }\bibfield  {title} {\enquote
  {\bibinfo {title} {Graphene bilayer with a twist: Electronic structure},}\
  }\href {https://link.aps.org/doi/10.1103/PhysRevLett.99.256802} {\bibfield
  {journal} {\bibinfo  {journal} {Phys. Rev. Lett.}\ }\textbf {\bibinfo
  {volume} {99}} (\bibinfo {year} {2007})}\BibitemShut {NoStop}%
\bibitem [{\citenamefont {Bistritzer}\ and\ \citenamefont
  {MacDonald}(2011)}]{Bistritzer2011}%
  \BibitemOpen
  \bibfield  {author} {\bibinfo {author} {\bibfnamefont {R.}~\bibnamefont
  {Bistritzer}}\ and\ \bibinfo {author} {\bibfnamefont {A.~H.}\ \bibnamefont
  {MacDonald}},\ }\bibfield  {title} {\enquote {\bibinfo {title} {Moire bands
  in twisted double-layer graphene},}\ }\href {\doibase
  10.1073/pnas.1108174108} {\bibfield  {journal} {\bibinfo  {journal} {PNAS}\
  }\textbf {\bibinfo {volume} {108}},\ \bibinfo {pages} {12233} (\bibinfo
  {year} {2011})}\BibitemShut {NoStop}%
\bibitem [{\citenamefont {Sharpe}\ \emph {et~al.}(2019)\citenamefont {Sharpe},
  \citenamefont {Fox}, \citenamefont {Barnard}, \citenamefont {Finney},
  \citenamefont {Watanabe}, \citenamefont {Taniguchi}, \citenamefont
  {Kastner},\ and\ \citenamefont {Goldhaber-Gordon}}]{Sharpe2019}%
  \BibitemOpen
  \bibfield  {author} {\bibinfo {author} {\bibfnamefont {A.~L.}\ \bibnamefont
  {Sharpe}}, \bibinfo {author} {\bibfnamefont {E.~J.}\ \bibnamefont {Fox}},
  \bibinfo {author} {\bibfnamefont {A.~W.}\ \bibnamefont {Barnard}}, \bibinfo
  {author} {\bibfnamefont {J.}~\bibnamefont {Finney}}, \bibinfo {author}
  {\bibfnamefont {K.}~\bibnamefont {Watanabe}}, \bibinfo {author}
  {\bibfnamefont {T.}~\bibnamefont {Taniguchi}}, \bibinfo {author}
  {\bibfnamefont {M.~A.}\ \bibnamefont {Kastner}}, \ and\ \bibinfo {author}
  {\bibfnamefont {D.}~\bibnamefont {Goldhaber-Gordon}},\ }\bibfield  {title}
  {\enquote {\bibinfo {title} {Emergent ferromagnetism near three-quarters
  filling in twisted bilayer graphene},}\ }\href {\doibase DOI:
  10.1126/science.aaw3780} {\bibfield  {journal} {\bibinfo  {journal}
  {Science}\ }\textbf {\bibinfo {volume} {365}},\ \bibinfo {pages} {605}
  (\bibinfo {year} {2019})}\BibitemShut {NoStop}%
\bibitem [{\citenamefont {Serlin}\ \emph {et~al.}(2019)\citenamefont {Serlin},
  \citenamefont {Tschirhart}, \citenamefont {Polshyn}, \citenamefont {Zhang},
  \citenamefont {Zhu}, \citenamefont {Watanabe}, \citenamefont {Taniguchi},
  \citenamefont {Balents},\ and\ \citenamefont {Young}}]{Serlin2019}%
  \BibitemOpen
  \bibfield  {author} {\bibinfo {author} {\bibfnamefont {M.}~\bibnamefont
  {Serlin}}, \bibinfo {author} {\bibfnamefont {C.~L.}\ \bibnamefont
  {Tschirhart}}, \bibinfo {author} {\bibfnamefont {H.}~\bibnamefont {Polshyn}},
  \bibinfo {author} {\bibfnamefont {Y.}~\bibnamefont {Zhang}}, \bibinfo
  {author} {\bibfnamefont {J.}~\bibnamefont {Zhu}}, \bibinfo {author}
  {\bibfnamefont {K.}~\bibnamefont {Watanabe}}, \bibinfo {author}
  {\bibfnamefont {T.}~\bibnamefont {Taniguchi}}, \bibinfo {author}
  {\bibfnamefont {L.}~\bibnamefont {Balents}}, \ and\ \bibinfo {author}
  {\bibfnamefont {A.~F.}\ \bibnamefont {Young}},\ }\bibfield  {title} {\enquote
  {\bibinfo {title} {Intrinsic quantized anomalous hall effect in a moir{\'{e}}
  heterostructure},}\ }\href {\doibase 10.1126/science.aay5533} {\bibfield
  {journal} {\bibinfo  {journal} {Science}\ }\textbf {\bibinfo {volume}
  {367}},\ \bibinfo {pages} {900} (\bibinfo {year} {2019})}\BibitemShut
  {NoStop}%
\bibitem [{\citenamefont {Kerelsky}\ \emph {et~al.}(2019)\citenamefont
  {Kerelsky}, \citenamefont {McGilly}, \citenamefont {Kennes}, \citenamefont
  {Xian}, \citenamefont {Yankowitz}, \citenamefont {Chen}, \citenamefont
  {Watanabe}, \citenamefont {Taniguchi}, \citenamefont {Hone}, \citenamefont
  {Dean}, \citenamefont {Rubio},\ and\ \citenamefont
  {Pasupathy}}]{Kerelsky2019}%
  \BibitemOpen
  \bibfield  {author} {\bibinfo {author} {\bibfnamefont {A.}~\bibnamefont
  {Kerelsky}}, \bibinfo {author} {\bibfnamefont {L.~J.}\ \bibnamefont
  {McGilly}}, \bibinfo {author} {\bibfnamefont {D.~M.}\ \bibnamefont {Kennes}},
  \bibinfo {author} {\bibfnamefont {L.}~\bibnamefont {Xian}}, \bibinfo {author}
  {\bibfnamefont {M.}~\bibnamefont {Yankowitz}}, \bibinfo {author}
  {\bibfnamefont {S.}~\bibnamefont {Chen}}, \bibinfo {author} {\bibfnamefont
  {K.}~\bibnamefont {Watanabe}}, \bibinfo {author} {\bibfnamefont
  {T.}~\bibnamefont {Taniguchi}}, \bibinfo {author} {\bibfnamefont
  {J.}~\bibnamefont {Hone}}, \bibinfo {author} {\bibfnamefont {C.}~\bibnamefont
  {Dean}}, \bibinfo {author} {\bibfnamefont {A.}~\bibnamefont {Rubio}}, \ and\
  \bibinfo {author} {\bibfnamefont {A.~N.}\ \bibnamefont {Pasupathy}},\
  }\bibfield  {title} {\enquote {\bibinfo {title} {Maximized electron
  interactions at the magic angle in twisted bilayer graphene},}\ }\href
  {https://doi.org/10.1038/s41586-019-1431-9} {\bibfield  {journal} {\bibinfo
  {journal} {Nature}\ }\textbf {\bibinfo {volume} {572}},\ \bibinfo {pages}
  {95} (\bibinfo {year} {2019})}\BibitemShut {NoStop}%
\bibitem [{\citenamefont {Xie}\ \emph {et~al.}(2019)\citenamefont {Xie},
  \citenamefont {Lian}, \citenamefont {Jäck}, \citenamefont {Liu},
  \citenamefont {Chiu}, \citenamefont {Watanabe}, \citenamefont {Taniguchi},
  \citenamefont {Bernevig},\ and\ \citenamefont {Yazdani}}]{Xie2019}%
  \BibitemOpen
  \bibfield  {author} {\bibinfo {author} {\bibfnamefont {Y.}~\bibnamefont
  {Xie}}, \bibinfo {author} {\bibfnamefont {B.}~\bibnamefont {Lian}}, \bibinfo
  {author} {\bibfnamefont {B.}~\bibnamefont {Jäck}}, \bibinfo {author}
  {\bibfnamefont {X.}~\bibnamefont {Liu}}, \bibinfo {author} {\bibfnamefont
  {C.-L.}\ \bibnamefont {Chiu}}, \bibinfo {author} {\bibfnamefont
  {K.}~\bibnamefont {Watanabe}}, \bibinfo {author} {\bibfnamefont
  {T.}~\bibnamefont {Taniguchi}}, \bibinfo {author} {\bibfnamefont {B.~A.}\
  \bibnamefont {Bernevig}}, \ and\ \bibinfo {author} {\bibfnamefont
  {A.}~\bibnamefont {Yazdani}},\ }\bibfield  {title} {\enquote {\bibinfo
  {title} {Spectroscopic signatures of many-body correlations in magic-angle
  twisted bilayer graphene},}\ }\href
  {https://doi.org/10.1038/s41586-019-1422-x} {\bibfield  {journal} {\bibinfo
  {journal} {Nature}\ }\textbf {\bibinfo {volume} {572}},\ \bibinfo {pages}
  {101} (\bibinfo {year} {2019})}\BibitemShut {NoStop}%
\bibitem [{\citenamefont {Jiang}\ \emph {et~al.}(2019)\citenamefont {Jiang},
  \citenamefont {Lai}, \citenamefont {Watanabe}, \citenamefont {Taniguchi},
  \citenamefont {Haule}, \citenamefont {Mao},\ and\ \citenamefont
  {Andrei}}]{Jiang2019}%
  \BibitemOpen
  \bibfield  {author} {\bibinfo {author} {\bibfnamefont {Yuhang}\ \bibnamefont
  {Jiang}}, \bibinfo {author} {\bibfnamefont {Xinyuan}\ \bibnamefont {Lai}},
  \bibinfo {author} {\bibfnamefont {Kenji}\ \bibnamefont {Watanabe}}, \bibinfo
  {author} {\bibfnamefont {Takashi}\ \bibnamefont {Taniguchi}}, \bibinfo
  {author} {\bibfnamefont {Kristjan}\ \bibnamefont {Haule}}, \bibinfo {author}
  {\bibfnamefont {Jinhai}\ \bibnamefont {Mao}}, \ and\ \bibinfo {author}
  {\bibfnamefont {Eva~Y.}\ \bibnamefont {Andrei}},\ }\bibfield  {title}
  {\enquote {\bibinfo {title} {Charge order and broken rotational symmetry in
  magic-angle twisted bilayer graphene},}\ }\href
  {https://doi.org/10.1038/s41586-019-1460-4} {\bibfield  {journal} {\bibinfo
  {journal} {Nature}\ }\textbf {\bibinfo {volume} {573}},\ \bibinfo {pages}
  {91} (\bibinfo {year} {2019})}\BibitemShut {NoStop}%
\bibitem [{\citenamefont {Bouhon}\ \emph {et~al.}(2019)\citenamefont {Bouhon},
  \citenamefont {Black-Schaffer},\ and\ \citenamefont {Slager}}]{Bouhon2019}%
  \BibitemOpen
  \bibfield  {author} {\bibinfo {author} {\bibfnamefont {A.}~\bibnamefont
  {Bouhon}}, \bibinfo {author} {\bibfnamefont {A.~M.}\ \bibnamefont
  {Black-Schaffer}}, \ and\ \bibinfo {author} {\bibfnamefont {R.-J.}\
  \bibnamefont {Slager}},\ }\bibfield  {title} {\enquote {\bibinfo {title}
  {Wilson loop approach to fragile topology of split elementary band
  representations and topological crystalline insulators with time-reversal
  symmetry},}\ }\href {https://link.aps.org/doi/10.1103/PhysRevB.100.195135}
  {\bibfield  {journal} {\bibinfo  {journal} {Phys. Rev. B}\ }\textbf {\bibinfo
  {volume} {100}},\ \bibinfo {pages} {195135} (\bibinfo {year}
  {2019})}\BibitemShut {NoStop}%
\bibitem [{\citenamefont {Ahn}\ \emph {et~al.}(2019)\citenamefont {Ahn},
  \citenamefont {Park},\ and\ \citenamefont {Yang}}]{Ahn2019}%
  \BibitemOpen
  \bibfield  {author} {\bibinfo {author} {\bibfnamefont {J.}~\bibnamefont
  {Ahn}}, \bibinfo {author} {\bibfnamefont {S.}~\bibnamefont {Park}}, \ and\
  \bibinfo {author} {\bibfnamefont {B.-J.}\ \bibnamefont {Yang}},\ }\bibfield
  {title} {\enquote {\bibinfo {title} {Failure of nielsen-ninomiya theorem and
  fragile topology in two-dimensional systems with space-time inversion
  symmetry: Application to twisted bilayer graphene at magic angle},}\ }\href
  {https://link.aps.org/doi/10.1103/PhysRevX.9.021013} {\bibfield  {journal}
  {\bibinfo  {journal} {Phys. Rev. X}\ }\textbf {\bibinfo {volume} {9}}
  (\bibinfo {year} {2019})}\BibitemShut {NoStop}%
\bibitem [{\citenamefont {Song}\ \emph {et~al.}(2019)\citenamefont {Song},
  \citenamefont {Wang}, \citenamefont {Shi}, \citenamefont {Li}, \citenamefont
  {Fang},\ and\ \citenamefont {Bernevig}}]{Song2019}%
  \BibitemOpen
  \bibfield  {author} {\bibinfo {author} {\bibfnamefont {Z.}~\bibnamefont
  {Song}}, \bibinfo {author} {\bibfnamefont {Z.}~\bibnamefont {Wang}}, \bibinfo
  {author} {\bibfnamefont {W.}~\bibnamefont {Shi}}, \bibinfo {author}
  {\bibfnamefont {G.}~\bibnamefont {Li}}, \bibinfo {author} {\bibfnamefont
  {C.}~\bibnamefont {Fang}}, \ and\ \bibinfo {author} {\bibfnamefont {B.~A.}\
  \bibnamefont {Bernevig}},\ }\bibfield  {title} {\enquote {\bibinfo {title}
  {All magic angles in twisted bilayer graphene are topological},}\ }\href
  {\doibase 10.1103/PhysRevLett.123.036401} {\bibfield  {journal} {\bibinfo
  {journal} {Phys. Rev. Lett.}\ }\textbf {\bibinfo {volume} {123}},\ \bibinfo
  {pages} {036401} (\bibinfo {year} {2019})}\BibitemShut {NoStop}%
\bibitem [{\citenamefont {Xie}\ \emph {et~al.}(2020)\citenamefont {Xie},
  \citenamefont {Song}, \citenamefont {Lian},\ and\ \citenamefont
  {Bernevig}}]{Xie2019a}%
  \BibitemOpen
  \bibfield  {author} {\bibinfo {author} {\bibfnamefont {F.}~\bibnamefont
  {Xie}}, \bibinfo {author} {\bibfnamefont {Z.}~\bibnamefont {Song}}, \bibinfo
  {author} {\bibfnamefont {B.}~\bibnamefont {Lian}}, \ and\ \bibinfo {author}
  {\bibfnamefont {B.~A.}\ \bibnamefont {Bernevig}},\ }\bibfield  {title}
  {\enquote {\bibinfo {title} {Topology-bounded superfluid weight in twisted
  bilayer graphene},}\ }\href
  {https://link.aps.org/doi/10.1103/PhysRevLett.124.167002} {\bibfield
  {journal} {\bibinfo  {journal} {Phys. Rev. Lett.}\ }\textbf {\bibinfo
  {volume} {124}},\ \bibinfo {pages} {167002} (\bibinfo {year}
  {2020})}\BibitemShut {NoStop}%
\bibitem [{\citenamefont {Lu}\ \emph {et~al.}(2021)\citenamefont {Lu},
  \citenamefont {Lian}, \citenamefont {Chaudhary}, \citenamefont {Piot},
  \citenamefont {G.}, \citenamefont {Watanabe}, \citenamefont {Taniguchi},
  \citenamefont {Poggio}, \citenamefont {MacDonald}, \citenamefont {Bernevig},\
  and\ \citenamefont {Efetov}}]{Lu2021}%
  \BibitemOpen
  \bibfield  {author} {\bibinfo {author} {\bibfnamefont {X.}~\bibnamefont
  {Lu}}, \bibinfo {author} {\bibfnamefont {B.}~\bibnamefont {Lian}}, \bibinfo
  {author} {\bibfnamefont {G.}~\bibnamefont {Chaudhary}}, \bibinfo {author}
  {\bibfnamefont {B.~A.}\ \bibnamefont {Piot}}, \bibinfo {author} {\bibnamefont
  {G.}}, \bibinfo {author} {\bibfnamefont {K.}~\bibnamefont {Watanabe}},
  \bibinfo {author} {\bibfnamefont {T.}~\bibnamefont {Taniguchi}}, \bibinfo
  {author} {\bibfnamefont {M.}~\bibnamefont {Poggio}}, \bibinfo {author}
  {\bibfnamefont {A.~H.}\ \bibnamefont {MacDonald}}, \bibinfo {author}
  {\bibfnamefont {B.~A.}\ \bibnamefont {Bernevig}}, \ and\ \bibinfo {author}
  {\bibfnamefont {D.~K.}\ \bibnamefont {Efetov}},\ }\bibfield  {title}
  {\enquote {\bibinfo {title} {Multiple flat bands and topological hofstadter
  butterfly in twisted bilayer graphene close to the second magic angle},}\
  }\href {\doibase 10.1073/pnas.2100006118} {\bibfield  {journal} {\bibinfo
  {journal} {PNAS}\ }\textbf {\bibinfo {volume} {118}},\ \bibinfo {pages}
  {e2100006118} (\bibinfo {year} {2021})}\BibitemShut {NoStop}%
\bibitem [{\citenamefont {X.}\ \emph {et~al.}(2019)\citenamefont {X.},
  \citenamefont {Stepanov}, \citenamefont {Yang}, \citenamefont {Xie},
  \citenamefont {Aamir}, \citenamefont {Das}, \citenamefont {Urgell},
  \citenamefont {Watanabe}, \citenamefont {Taniguchi}, \citenamefont {Zhang},
  \citenamefont {Bachtold}, \citenamefont {MacDonald},\ and\ \citenamefont
  {Efetov}}]{Lu2019}%
  \BibitemOpen
  \bibfield  {author} {\bibinfo {author} {\bibnamefont {X.}}, \bibinfo {author}
  {\bibfnamefont {P.}~\bibnamefont {Stepanov}}, \bibinfo {author}
  {\bibfnamefont {W.}~\bibnamefont {Yang}}, \bibinfo {author} {\bibfnamefont
  {M.}~\bibnamefont {Xie}}, \bibinfo {author} {\bibfnamefont {M.~A.}\
  \bibnamefont {Aamir}}, \bibinfo {author} {\bibfnamefont {I.}~\bibnamefont
  {Das}}, \bibinfo {author} {\bibfnamefont {C.}~\bibnamefont {Urgell}},
  \bibinfo {author} {\bibfnamefont {K.}~\bibnamefont {Watanabe}}, \bibinfo
  {author} {\bibfnamefont {T.}~\bibnamefont {Taniguchi}}, \bibinfo {author}
  {\bibfnamefont {G.}~\bibnamefont {Zhang}}, \bibinfo {author} {\bibfnamefont
  {A.}~\bibnamefont {Bachtold}}, \bibinfo {author} {\bibfnamefont {A.~H.}\
  \bibnamefont {MacDonald}}, \ and\ \bibinfo {author} {\bibfnamefont {D.~K.}\
  \bibnamefont {Efetov}},\ }\bibfield  {title} {\enquote {\bibinfo {title}
  {Superconductors, orbital magnets and correlated states in magic-angle
  bilayer graphene},}\ }\href {\doibase 10.1038/s41586-019-1695-0} {\bibfield
  {journal} {\bibinfo  {journal} {Nature}\ }\textbf {\bibinfo {volume} {574}},\
  \bibinfo {pages} {653--657} (\bibinfo {year} {2019})}\BibitemShut {NoStop}%
\bibitem [{\citenamefont {Wu}\ \emph {et~al.}(2018)\citenamefont {Wu},
  \citenamefont {Lovorn}, \citenamefont {Tutuc},\ and\ \citenamefont
  {MacDonald}}]{Wu2018}%
  \BibitemOpen
  \bibfield  {author} {\bibinfo {author} {\bibfnamefont {F.}~\bibnamefont
  {Wu}}, \bibinfo {author} {\bibfnamefont {T.}~\bibnamefont {Lovorn}}, \bibinfo
  {author} {\bibfnamefont {E.}~\bibnamefont {Tutuc}}, \ and\ \bibinfo {author}
  {\bibfnamefont {A.~H.}\ \bibnamefont {MacDonald}},\ }\bibfield  {title}
  {\enquote {\bibinfo {title} {Hubbard model physics in transition metal
  dichalcogenide moir\'e bands},}\ }\href
  {https://link.aps.org/doi/10.1103/PhysRevLett.121.026402} {\bibfield
  {journal} {\bibinfo  {journal} {Phys. Rev. Lett.}\ }\textbf {\bibinfo
  {volume} {121}},\ \bibinfo {pages} {026402} (\bibinfo {year}
  {2018})}\BibitemShut {NoStop}%
\bibitem [{\citenamefont {Wu}\ \emph {et~al.}(2019)\citenamefont {Wu},
  \citenamefont {Lovorn}, \citenamefont {Tutuc}, \citenamefont {Martin},\ and\
  \citenamefont {MacDonald}}]{Wu2019}%
  \BibitemOpen
  \bibfield  {author} {\bibinfo {author} {\bibfnamefont {F.}~\bibnamefont
  {Wu}}, \bibinfo {author} {\bibfnamefont {T.}~\bibnamefont {Lovorn}}, \bibinfo
  {author} {\bibfnamefont {E.}~\bibnamefont {Tutuc}}, \bibinfo {author}
  {\bibfnamefont {I.}~\bibnamefont {Martin}}, \ and\ \bibinfo {author}
  {\bibfnamefont {A.~H.}\ \bibnamefont {MacDonald}},\ }\bibfield  {title}
  {\enquote {\bibinfo {title} {Topological insulators in twisted transition
  metal dichalcogenide homobilayers},}\ }\href
  {https://link.aps.org/doi/10.1103/PhysRevLett.122.086402} {\bibfield
  {journal} {\bibinfo  {journal} {Phys. Rev. Lett.}\ }\textbf {\bibinfo
  {volume} {122}},\ \bibinfo {pages} {086402} (\bibinfo {year}
  {2019})}\BibitemShut {NoStop}%
\bibitem [{\citenamefont {Tang}\ \emph {et~al.}(2020)\citenamefont {Tang},
  \citenamefont {Li}, \citenamefont {Li}, \citenamefont {Xu}, \citenamefont
  {Liu}, \citenamefont {Barmak}, \citenamefont {Watanabe}, \citenamefont
  {Taniguchi}, \citenamefont {MacDonald}, \citenamefont {Shan},\ and\
  \citenamefont {Mak}}]{Tang2020}%
  \BibitemOpen
  \bibfield  {author} {\bibinfo {author} {\bibfnamefont {Y.}~\bibnamefont
  {Tang}}, \bibinfo {author} {\bibfnamefont {L.}~\bibnamefont {Li}}, \bibinfo
  {author} {\bibfnamefont {T.}~\bibnamefont {Li}}, \bibinfo {author}
  {\bibfnamefont {Y.}~\bibnamefont {Xu}}, \bibinfo {author} {\bibfnamefont
  {S.}~\bibnamefont {Liu}}, \bibinfo {author} {\bibfnamefont {K.}~\bibnamefont
  {Barmak}}, \bibinfo {author} {\bibfnamefont {K.}~\bibnamefont {Watanabe}},
  \bibinfo {author} {\bibfnamefont {T.}~\bibnamefont {Taniguchi}}, \bibinfo
  {author} {\bibfnamefont {A.~H.}\ \bibnamefont {MacDonald}}, \bibinfo {author}
  {\bibfnamefont {J.}~\bibnamefont {Shan}}, \ and\ \bibinfo {author}
  {\bibfnamefont {K.~F.}\ \bibnamefont {Mak}},\ }\bibfield  {title} {\enquote
  {\bibinfo {title} {Simulation of hubbard model physics in wse2/ws2
  moirésuperlattices},}\ }\href {https://doi.org/10.1038/s41586-020-2085-3}
  {\bibfield  {journal} {\bibinfo  {journal} {Nature}\ }\textbf {\bibinfo
  {volume} {579}},\ \bibinfo {pages} {353} (\bibinfo {year}
  {2020})}\BibitemShut {NoStop}%
\bibitem [{\citenamefont {Regan}\ \emph {et~al.}(2020)\citenamefont {Regan},
  \citenamefont {Wang}, \citenamefont {Jin}, \citenamefont {Bakti},
  \citenamefont {Gao}, \citenamefont {Wei}, \citenamefont {Zhao}, \citenamefont
  {Zhao}, \citenamefont {Zhang}, \citenamefont {Yumigeta}, \citenamefont
  {Blei}, \citenamefont {Carlström}, \citenamefont {Watanabe}, \citenamefont
  {Taniguchi}, \citenamefont {Tongay}, \citenamefont {Crommie}, \citenamefont
  {Zettl},\ and\ \citenamefont {Wang}}]{Regan2020}%
  \BibitemOpen
  \bibfield  {author} {\bibinfo {author} {\bibfnamefont {E.~C.}\ \bibnamefont
  {Regan}}, \bibinfo {author} {\bibfnamefont {D.}~\bibnamefont {Wang}},
  \bibinfo {author} {\bibfnamefont {C.}~\bibnamefont {Jin}}, \bibinfo {author}
  {\bibfnamefont {M.~I.~Utama}\ \bibnamefont {Bakti}}, \bibinfo {author}
  {\bibfnamefont {B.}~\bibnamefont {Gao}}, \bibinfo {author} {\bibfnamefont
  {X.}~\bibnamefont {Wei}}, \bibinfo {author} {\bibfnamefont {S.}~\bibnamefont
  {Zhao}}, \bibinfo {author} {\bibfnamefont {W.}~\bibnamefont {Zhao}}, \bibinfo
  {author} {\bibfnamefont {Z.}~\bibnamefont {Zhang}}, \bibinfo {author}
  {\bibfnamefont {K.}~\bibnamefont {Yumigeta}}, \bibinfo {author}
  {\bibfnamefont {M.}~\bibnamefont {Blei}}, \bibinfo {author} {\bibfnamefont
  {J.~D.}\ \bibnamefont {Carlström}}, \bibinfo {author} {\bibfnamefont
  {K.}~\bibnamefont {Watanabe}}, \bibinfo {author} {\bibfnamefont
  {T.}~\bibnamefont {Taniguchi}}, \bibinfo {author} {\bibfnamefont
  {S.}~\bibnamefont {Tongay}}, \bibinfo {author} {\bibfnamefont
  {M.}~\bibnamefont {Crommie}}, \bibinfo {author} {\bibfnamefont
  {A.}~\bibnamefont {Zettl}}, \ and\ \bibinfo {author} {\bibfnamefont
  {F.}~\bibnamefont {Wang}},\ }\bibfield  {title} {\enquote {\bibinfo {title}
  {Mott and generalized wigner crystal states in wse2/ws2 moiré
  superlattices},}\ }\href {https://doi.org/10.1038/s41586-020-2092-4}
  {\bibfield  {journal} {\bibinfo  {journal} {Nature}\ }\textbf {\bibinfo
  {volume} {579}},\ \bibinfo {pages} {359} (\bibinfo {year}
  {2020})}\BibitemShut {NoStop}%
\bibitem [{\citenamefont {Xu}\ \emph {et~al.}(2020)\citenamefont {Xu},
  \citenamefont {Liu}, \citenamefont {Rhodes}, \citenamefont {Watanabe},
  \citenamefont {Taniguchi}, \citenamefont {Hone}, \citenamefont {Elser},
  \citenamefont {Mak},\ and\ \citenamefont {Shan}}]{Xu2020}%
  \BibitemOpen
  \bibfield  {author} {\bibinfo {author} {\bibfnamefont {Y.}~\bibnamefont
  {Xu}}, \bibinfo {author} {\bibfnamefont {S.}~\bibnamefont {Liu}}, \bibinfo
  {author} {\bibfnamefont {D.~A.}\ \bibnamefont {Rhodes}}, \bibinfo {author}
  {\bibfnamefont {K.}~\bibnamefont {Watanabe}}, \bibinfo {author}
  {\bibfnamefont {T.}~\bibnamefont {Taniguchi}}, \bibinfo {author}
  {\bibfnamefont {J.}~\bibnamefont {Hone}}, \bibinfo {author} {\bibfnamefont
  {V.}~\bibnamefont {Elser}}, \bibinfo {author} {\bibfnamefont {K.~F.}\
  \bibnamefont {Mak}}, \ and\ \bibinfo {author} {\bibfnamefont
  {J.}~\bibnamefont {Shan}},\ }\bibfield  {title} {\enquote {\bibinfo {title}
  {Correlated insulating states at fractional fillings of
  moirésuperlattices},}\ }\href {https://doi.org/10.1038/s41586-020-2868-6}
  {\bibfield  {journal} {\bibinfo  {journal} {Nature}\ }\textbf {\bibinfo
  {volume} {587}},\ \bibinfo {pages} {214} (\bibinfo {year}
  {2020})}\BibitemShut {NoStop}%
\bibitem [{\citenamefont {Zhang}\ \emph {et~al.}(2021)\citenamefont {Zhang},
  \citenamefont {Wang}, \citenamefont {Watanabe}, \citenamefont {Taniguchi},
  \citenamefont {Ueno}, \citenamefont {Tutuc},\ and\ \citenamefont
  {LeRoy}}]{Zhang2020}%
  \BibitemOpen
  \bibfield  {author} {\bibinfo {author} {\bibfnamefont {Z.}~\bibnamefont
  {Zhang}}, \bibinfo {author} {\bibfnamefont {Y.}~\bibnamefont {Wang}},
  \bibinfo {author} {\bibfnamefont {K.}~\bibnamefont {Watanabe}}, \bibinfo
  {author} {\bibfnamefont {T.}~\bibnamefont {Taniguchi}}, \bibinfo {author}
  {\bibfnamefont {K.}~\bibnamefont {Ueno}}, \bibinfo {author} {\bibfnamefont
  {E.}~\bibnamefont {Tutuc}}, \ and\ \bibinfo {author} {\bibfnamefont {B.~J.}\
  \bibnamefont {LeRoy}},\ }\bibfield  {title} {\enquote {\bibinfo {title} {Flat
  bands in twisted bilayer transition metal dichalcogenides},}\ }\href
  {https://doi.org/10.1038/s41567-020-0958-x} {\bibfield  {journal} {\bibinfo
  {journal} {Nat. Phys.}\ }\textbf {\bibinfo {volume} {16}},\ \bibinfo {pages}
  {1093} (\bibinfo {year} {2021})}\BibitemShut {NoStop}%
\bibitem [{\citenamefont {Slagle}\ and\ \citenamefont {Fu}(2020)}]{Slagle2020}%
  \BibitemOpen
  \bibfield  {author} {\bibinfo {author} {\bibfnamefont {K.}~\bibnamefont
  {Slagle}}\ and\ \bibinfo {author} {\bibfnamefont {L.}~\bibnamefont {Fu}},\
  }\bibfield  {title} {\enquote {\bibinfo {title} {Charge transfer excitations,
  pair density waves, and superconductivity in moir\'e materials},}\ }\href
  {https://link.aps.org/doi/10.1103/PhysRevB.102.235423} {\bibfield  {journal}
  {\bibinfo  {journal} {Phys. Rev. B}\ }\textbf {\bibinfo {volume} {102}},\
  \bibinfo {pages} {235423} (\bibinfo {year} {2020})}\BibitemShut {NoStop}%
\bibitem [{\citenamefont {Bi}\ and\ \citenamefont {Fu}(2021)}]{Bi2021}%
  \BibitemOpen
  \bibfield  {author} {\bibinfo {author} {\bibfnamefont {Z.}~\bibnamefont
  {Bi}}\ and\ \bibinfo {author} {\bibfnamefont {L.}~\bibnamefont {Fu}},\
  }\bibfield  {title} {\enquote {\bibinfo {title} {Excitonic density wave and
  spin-valley superfluid in bilayer transition metal dichalcogenide},}\ }\href
  {https://doi.org/10.1038/s41467-020-20802-z} {\bibfield  {journal} {\bibinfo
  {journal} {Nat. Commun.}\ }\textbf {\bibinfo {volume} {12}},\ \bibinfo
  {pages} {642} (\bibinfo {year} {2021})}\BibitemShut {NoStop}%
\bibitem [{\citenamefont {Morales-Dur\'an}\ \emph {et~al.}(2021)\citenamefont
  {Morales-Dur\'an}, \citenamefont {MacDonald},\ and\ \citenamefont
  {Potasz}}]{Duran2021}%
  \BibitemOpen
  \bibfield  {author} {\bibinfo {author} {\bibfnamefont {N.}~\bibnamefont
  {Morales-Dur\'an}}, \bibinfo {author} {\bibfnamefont {A.~H.}\ \bibnamefont
  {MacDonald}}, \ and\ \bibinfo {author} {\bibfnamefont {P.}~\bibnamefont
  {Potasz}},\ }\bibfield  {title} {\enquote {\bibinfo {title} {Metal-insulator
  transition in transition metal dichalcogenide heterobilayer moir\'e
  superlattices},}\ }\href
  {https://link.aps.org/doi/10.1103/PhysRevB.103.L241110} {\bibfield  {journal}
  {\bibinfo  {journal} {Phys. Rev. B}\ }\textbf {\bibinfo {volume} {103}},\
  \bibinfo {pages} {L241110} (\bibinfo {year} {2021})}\BibitemShut {NoStop}%
\bibitem [{\citenamefont {Lian}\ \emph {et~al.}(2020)\citenamefont {Lian},
  \citenamefont {Liu}, \citenamefont {Zhang},\ and\ \citenamefont
  {Wang}}]{Lian2020}%
  \BibitemOpen
  \bibfield  {author} {\bibinfo {author} {\bibfnamefont {B.}~\bibnamefont
  {Lian}}, \bibinfo {author} {\bibfnamefont {Z.}~\bibnamefont {Liu}}, \bibinfo
  {author} {\bibfnamefont {Y.}~\bibnamefont {Zhang}}, \ and\ \bibinfo {author}
  {\bibfnamefont {J.}~\bibnamefont {Wang}},\ }\bibfield  {title} {\enquote
  {\bibinfo {title} {Flat chern band from twisted bilayer
  ${\mathrm{mnbi}}_{2}{\mathrm{te}}_{4}$},}\ }\href {\doibase
  10.1103/PhysRevLett.124.126402} {\bibfield  {journal} {\bibinfo  {journal}
  {Phys. Rev. Lett.}\ }\textbf {\bibinfo {volume} {124}},\ \bibinfo {pages}
  {126402} (\bibinfo {year} {2020})}\BibitemShut {NoStop}%
\bibitem [{\citenamefont {Liu}\ \emph {et~al.}()\citenamefont {Liu},
  \citenamefont {Wang},\ and\ \citenamefont {Wang}}]{Liu2021}%
  \BibitemOpen
  \bibfield  {author} {\bibinfo {author} {\bibfnamefont {Z.}~\bibnamefont
  {Liu}}, \bibinfo {author} {\bibfnamefont {H.}~\bibnamefont {Wang}}, \ and\
  \bibinfo {author} {\bibfnamefont {J.}~\bibnamefont {Wang}},\ }\bibfield
  {title} {\enquote {\bibinfo {title} {Magnetic moiré surface states and flat
  chern band in topological insulators},}\ }\href@noop {} {\ }\Eprint
  {http://arxiv.org/abs/arXiv:2106.01630} {arXiv:2106.01630} \BibitemShut
  {NoStop}%
\bibitem [{\citenamefont {Cano}\ \emph {et~al.}(2021)\citenamefont {Cano},
  \citenamefont {Fang}, \citenamefont {Pixley},\ and\ \citenamefont
  {Wilson}}]{Cano2021}%
  \BibitemOpen
  \bibfield  {author} {\bibinfo {author} {\bibfnamefont {J.}~\bibnamefont
  {Cano}}, \bibinfo {author} {\bibfnamefont {S.}~\bibnamefont {Fang}}, \bibinfo
  {author} {\bibfnamefont {J.~H.}\ \bibnamefont {Pixley}}, \ and\ \bibinfo
  {author} {\bibfnamefont {J.~H.}\ \bibnamefont {Wilson}},\ }\bibfield  {title}
  {\enquote {\bibinfo {title} {Moir\'e superlattice on the surface of a
  topological insulator},}\ }\href
  {https://link.aps.org/doi/10.1103/PhysRevB.103.155157} {\bibfield  {journal}
  {\bibinfo  {journal} {Phys. Rev. B}\ }\textbf {\bibinfo {volume} {103}},\
  \bibinfo {pages} {155157} (\bibinfo {year} {2021})}\BibitemShut {NoStop}%
\bibitem [{\citenamefont {Wang}\ \emph {et~al.}(2021)\citenamefont {Wang},
  \citenamefont {Yuan},\ and\ \citenamefont {Fu}}]{Wang2021}%
  \BibitemOpen
  \bibfield  {author} {\bibinfo {author} {\bibfnamefont {T.}~\bibnamefont
  {Wang}}, \bibinfo {author} {\bibfnamefont {N.~F.~Q.}\ \bibnamefont {Yuan}}, \
  and\ \bibinfo {author} {\bibfnamefont {L.}~\bibnamefont {Fu}},\ }\bibfield
  {title} {\enquote {\bibinfo {title} {Moir\'e surface states and enhanced
  superconductivity in topological insulators},}\ }\href {\doibase
  10.1103/PhysRevX.11.021024} {\bibfield  {journal} {\bibinfo  {journal} {Phys.
  Rev. X}\ }\textbf {\bibinfo {volume} {11}},\ \bibinfo {pages} {021024}
  (\bibinfo {year} {2021})}\BibitemShut {NoStop}%
\bibitem [{\citenamefont {Dunbrack}\ and\ \citenamefont
  {Cano}()}]{Dunbrack2021}%
  \BibitemOpen
  \bibfield  {author} {\bibinfo {author} {\bibfnamefont {A.}~\bibnamefont
  {Dunbrack}}\ and\ \bibinfo {author} {\bibfnamefont {J.}~\bibnamefont
  {Cano}},\ }\bibfield  {title} {\enquote {\bibinfo {title} {Magic angle
  conditions for twisted 3d topological insulators},}\ }\href@noop {} {\
  }\Eprint {http://arxiv.org/abs/arXiv:2112.11464} {arXiv:2112.11464}
  \BibitemShut {NoStop}%
\bibitem [{\citenamefont {Volkov}\ \emph {et~al.}()\citenamefont {Volkov},
  \citenamefont {Wilson},\ and\ \citenamefont {Pixley}}]{Volkov2020}%
  \BibitemOpen
  \bibfield  {author} {\bibinfo {author} {\bibfnamefont {P.~A.}\ \bibnamefont
  {Volkov}}, \bibinfo {author} {\bibfnamefont {J.~H.}\ \bibnamefont {Wilson}},
  \ and\ \bibinfo {author} {\bibfnamefont {J.~H.}\ \bibnamefont {Pixley}},\
  }\bibfield  {title} {\enquote {\bibinfo {title} {Magic angles and
  current-induced topology in twisted nodal superconductors},}\ }\href@noop {}
  {\ }\Eprint {http://arxiv.org/abs/arXiv:2012.07860} {arXiv:2012.07860}
  \BibitemShut {NoStop}%
\bibitem [{\citenamefont {Can}\ \emph {et~al.}(2021)\citenamefont {Can},
  \citenamefont {Tummuru}, \citenamefont {Day}, \citenamefont {Elfimov},
  \citenamefont {Damascelli},\ and\ \citenamefont {Franz}}]{Can2021}%
  \BibitemOpen
  \bibfield  {author} {\bibinfo {author} {\bibfnamefont {O.}~\bibnamefont
  {Can}}, \bibinfo {author} {\bibfnamefont {T.}~\bibnamefont {Tummuru}},
  \bibinfo {author} {\bibfnamefont {R.~P.}\ \bibnamefont {Day}}, \bibinfo
  {author} {\bibfnamefont {I.}~\bibnamefont {Elfimov}}, \bibinfo {author}
  {\bibfnamefont {A.}~\bibnamefont {Damascelli}}, \ and\ \bibinfo {author}
  {\bibfnamefont {M.}~\bibnamefont {Franz}},\ }\bibfield  {title} {\enquote
  {\bibinfo {title} {High-temperature topological superconductivity in twisted
  double-layer copper oxides},}\ }\href {\doibase 10.1038/s41567-020-01142-7}
  {\bibfield  {journal} {\bibinfo  {journal} {Nat. Phys.}\ }\textbf {\bibinfo
  {volume} {17}},\ \bibinfo {pages} {519} (\bibinfo {year} {2021})}\BibitemShut
  {NoStop}%
\bibitem [{\citenamefont {Zhao}\ \emph {et~al.}()\citenamefont {Zhao},
  \citenamefont {Poccia}, \citenamefont {Cui}, \citenamefont {Volkov},
  \citenamefont {Yoo}, \citenamefont {Engelke}, \citenamefont {Ronen},
  \citenamefont {Zhong}, \citenamefont {Gu}, \citenamefont {Plugge},
  \citenamefont {Tummuru}, \citenamefont {Franz}, \citenamefont {Pixley},\ and\
  \citenamefont {Kim}}]{Zhao2021}%
  \BibitemOpen
  \bibfield  {author} {\bibinfo {author} {\bibfnamefont {S.~Y.~F.}\
  \bibnamefont {Zhao}}, \bibinfo {author} {\bibfnamefont {N.}~\bibnamefont
  {Poccia}}, \bibinfo {author} {\bibfnamefont {X.}~\bibnamefont {Cui}},
  \bibinfo {author} {\bibfnamefont {P.~A.}\ \bibnamefont {Volkov}}, \bibinfo
  {author} {\bibfnamefont {H.}~\bibnamefont {Yoo}}, \bibinfo {author}
  {\bibfnamefont {R.}~\bibnamefont {Engelke}}, \bibinfo {author} {\bibfnamefont
  {Y.}~\bibnamefont {Ronen}}, \bibinfo {author} {\bibfnamefont
  {R.}~\bibnamefont {Zhong}}, \bibinfo {author} {\bibfnamefont
  {G.}~\bibnamefont {Gu}}, \bibinfo {author} {\bibfnamefont {S.}~\bibnamefont
  {Plugge}}, \bibinfo {author} {\bibfnamefont {T.}~\bibnamefont {Tummuru}},
  \bibinfo {author} {\bibfnamefont {M.}~\bibnamefont {Franz}}, \bibinfo
  {author} {\bibfnamefont {J.~H.}\ \bibnamefont {Pixley}}, \ and\ \bibinfo
  {author} {\bibfnamefont {P.}~\bibnamefont {Kim}},\ }\bibfield  {title}
  {\enquote {\bibinfo {title} {Emergent interfacial superconductivity between
  twisted cuprate superconductors},}\ }\href@noop {} {\ }\Eprint
  {http://arxiv.org/abs/arXiv:2108.13455} {arXiv:2108.13455} \BibitemShut
  {NoStop}%
\bibitem [{\citenamefont {Burkov}(2018)}]{Burkov2018}%
  \BibitemOpen
  \bibfield  {author} {\bibinfo {author} {\bibfnamefont {A.~A.}\ \bibnamefont
  {Burkov}},\ }\bibfield  {title} {\enquote {\bibinfo {title} {Quantum
  anomalies in nodal line semimetals},}\ }\href {\doibase
  10.1103/PhysRevB.97.165104} {\bibfield  {journal} {\bibinfo  {journal} {Phys.
  Rev. B}\ }\textbf {\bibinfo {volume} {97}},\ \bibinfo {pages} {165104}
  (\bibinfo {year} {2018})}\BibitemShut {NoStop}%
\bibitem [{\citenamefont {Wilczek}\ and\ \citenamefont
  {Zee}(1984)}]{Wilczek1984}%
  \BibitemOpen
  \bibfield  {author} {\bibinfo {author} {\bibfnamefont {F.}~\bibnamefont
  {Wilczek}}\ and\ \bibinfo {author} {\bibfnamefont {A.}~\bibnamefont {Zee}},\
  }\bibfield  {title} {\enquote {\bibinfo {title} {Appearance of gauge
  structure in simple dynamical systems},}\ }\href
  {https://link.aps.org/doi/10.1103/PhysRevLett.52.2111} {\bibfield  {journal}
  {\bibinfo  {journal} {Phys. Rev. Lett.}\ }\textbf {\bibinfo {volume} {52}},\
  \bibinfo {pages} {2111} (\bibinfo {year} {1984})}\BibitemShut {NoStop}%
\bibitem [{\citenamefont {S.-Jose}\ \emph {et~al.}(2012)\citenamefont
  {S.-Jose}, \citenamefont {Gonz{\'{a}}lez},\ and\ \citenamefont
  {Guinea}}]{San-Jose2012}%
  \BibitemOpen
  \bibfield  {author} {\bibinfo {author} {\bibfnamefont {P.}~\bibnamefont
  {S.-Jose}}, \bibinfo {author} {\bibfnamefont {J.}~\bibnamefont
  {Gonz{\'{a}}lez}}, \ and\ \bibinfo {author} {\bibfnamefont {F.}~\bibnamefont
  {Guinea}},\ }\bibfield  {title} {\enquote {\bibinfo {title} {Non-abelian
  gauge potentials in graphene bilayers},}\ }\href
  {https://link.aps.org/doi/10.1103/PhysRevLett.108.216802} {\bibfield
  {journal} {\bibinfo  {journal} {Phys. Rev. Lett.}\ }\textbf {\bibinfo
  {volume} {108}} (\bibinfo {year} {2012})}\BibitemShut {NoStop}%
\bibitem [{\citenamefont {Tarnopolsky}\ \emph {et~al.}(2019)\citenamefont
  {Tarnopolsky}, \citenamefont {Kruchkov},\ and\ \citenamefont
  {Vishwanath}}]{Tarnopolsky2019}%
  \BibitemOpen
  \bibfield  {author} {\bibinfo {author} {\bibfnamefont {G.}~\bibnamefont
  {Tarnopolsky}}, \bibinfo {author} {\bibfnamefont {A.~J.}\ \bibnamefont
  {Kruchkov}}, \ and\ \bibinfo {author} {\bibfnamefont {A.}~\bibnamefont
  {Vishwanath}},\ }\bibfield  {title} {\enquote {\bibinfo {title} {Origin of
  magic angles in twisted bilayer graphene},}\ }\href
  {https://link.aps.org/doi/10.1103/PhysRevLett.122.106405} {\bibfield
  {journal} {\bibinfo  {journal} {Phys. Rev. Lett.}\ }\textbf {\bibinfo
  {volume} {122}} (\bibinfo {year} {2019})}\BibitemShut {NoStop}%
\bibitem [{\citenamefont {Guinea}\ \emph {et~al.}(2008)\citenamefont {Guinea},
  \citenamefont {Katsnelson},\ and\ \citenamefont {Vozmediano}}]{Guinea2008}%
  \BibitemOpen
  \bibfield  {author} {\bibinfo {author} {\bibfnamefont {F.}~\bibnamefont
  {Guinea}}, \bibinfo {author} {\bibfnamefont {M.~I.}\ \bibnamefont
  {Katsnelson}}, \ and\ \bibinfo {author} {\bibfnamefont {M.~A.~H.}\
  \bibnamefont {Vozmediano}},\ }\bibfield  {title} {\enquote {\bibinfo {title}
  {Midgap states and charge inhomogeneities in corrugated graphene},}\ }\href
  {https://link.aps.org/doi/10.1103/PhysRevB.77.075422} {\bibfield  {journal}
  {\bibinfo  {journal} {Phys. Rev. B}\ }\textbf {\bibinfo {volume} {77}},\
  \bibinfo {pages} {075422} (\bibinfo {year} {2008})}\BibitemShut {NoStop}%
\bibitem [{\citenamefont {Guinea}\ \emph {et~al.}(2010)\citenamefont {Guinea},
  \citenamefont {Katsnelson},\ and\ \citenamefont {Geim}}]{Guinea2010}%
  \BibitemOpen
  \bibfield  {author} {\bibinfo {author} {\bibfnamefont {F.}~\bibnamefont
  {Guinea}}, \bibinfo {author} {\bibfnamefont {M.~I.}\ \bibnamefont
  {Katsnelson}}, \ and\ \bibinfo {author} {\bibfnamefont {A.~K.}\ \bibnamefont
  {Geim}},\ }\bibfield  {title} {\enquote {\bibinfo {title} {Energy gaps and a
  zero-field quantum hall effect in graphene by strain engineering},}\ }\href
  {https://doi.org/10.1038/nphys1420} {\bibfield  {journal} {\bibinfo
  {journal} {Nat. Phys.}\ }\textbf {\bibinfo {volume} {6}},\ \bibinfo {pages}
  {30} (\bibinfo {year} {2010})}\BibitemShut {NoStop}%
\bibitem [{\citenamefont {Snyman}(2009)}]{Snyman2009}%
  \BibitemOpen
  \bibfield  {author} {\bibinfo {author} {\bibfnamefont {I.}~\bibnamefont
  {Snyman}},\ }\bibfield  {title} {\enquote {\bibinfo {title} {Gapped state of
  a carbon monolayer in periodic magnetic and electric fields},}\ }\href
  {https://link.aps.org/doi/10.1103/PhysRevB.80.054303} {\bibfield  {journal}
  {\bibinfo  {journal} {Phys. Rev. B}\ }\textbf {\bibinfo {volume} {80}},\
  \bibinfo {pages} {054303} (\bibinfo {year} {2009})}\BibitemShut {NoStop}%
\bibitem [{\citenamefont {Tang}\ and\ \citenamefont {Fu}(2014)}]{Tang2014}%
  \BibitemOpen
  \bibfield  {author} {\bibinfo {author} {\bibfnamefont {E.}~\bibnamefont
  {Tang}}\ and\ \bibinfo {author} {\bibfnamefont {L.}~\bibnamefont {Fu}},\
  }\bibfield  {title} {\enquote {\bibinfo {title} {Strain-induced partially
  flat band, helical snake states and interface superconductivity in
  topological crystalline insulators},}\ }\href {\doibase 10.1038/nphys3109}
  {\bibfield  {journal} {\bibinfo  {journal} {Nat. Phys}\ }\textbf {\bibinfo
  {volume} {10}},\ \bibinfo {pages} {964} (\bibinfo {year} {2014})}\BibitemShut
  {NoStop}%
\bibitem [{\citenamefont {Aharonov}\ and\ \citenamefont
  {Casher}(1979)}]{Aharonov1979}%
  \BibitemOpen
  \bibfield  {author} {\bibinfo {author} {\bibfnamefont {Y.}~\bibnamefont
  {Aharonov}}\ and\ \bibinfo {author} {\bibfnamefont {A.}~\bibnamefont
  {Casher}},\ }\bibfield  {title} {\enquote {\bibinfo {title} {Ground state of
  a spin-\textonehalf{} charged particle in a two-dimensional magnetic
  field},}\ }\href {https://link.aps.org/doi/10.1103/PhysRevA.19.2461}
  {\bibfield  {journal} {\bibinfo  {journal} {Phys. Rev. A}\ }\textbf {\bibinfo
  {volume} {19}},\ \bibinfo {pages} {2461} (\bibinfo {year}
  {1979})}\BibitemShut {NoStop}%
\bibitem [{\citenamefont {Liu}\ \emph {et~al.}(2019)\citenamefont {Liu},
  \citenamefont {Liu},\ and\ \citenamefont {Dai}}]{Liu2019}%
  \BibitemOpen
  \bibfield  {author} {\bibinfo {author} {\bibfnamefont {J.}~\bibnamefont
  {Liu}}, \bibinfo {author} {\bibfnamefont {J.}~\bibnamefont {Liu}}, \ and\
  \bibinfo {author} {\bibfnamefont {X.}~\bibnamefont {Dai}},\ }\bibfield
  {title} {\enquote {\bibinfo {title} {Pseudo landau level representation of
  twisted bilayer graphene: Band topology and implications on the correlated
  insulating phase},}\ }\href
  {https://link.aps.org/doi/10.1103/PhysRevB.99.155415} {\bibfield  {journal}
  {\bibinfo  {journal} {Phys. Rev. B}\ }\textbf {\bibinfo {volume} {99}},\
  \bibinfo {pages} {155415} (\bibinfo {year} {2019})}\BibitemShut {NoStop}%
\bibitem [{\citenamefont {Luo}\ \emph {et~al.}(2021)\citenamefont {Luo},
  \citenamefont {Xu},\ and\ \citenamefont {Jian}}]{Luo2021}%
  \BibitemOpen
  \bibfield  {author} {\bibinfo {author} {\bibfnamefont {Z.-X.}\ \bibnamefont
  {Luo}}, \bibinfo {author} {\bibfnamefont {C.}~\bibnamefont {Xu}}, \ and\
  \bibinfo {author} {\bibfnamefont {C.-M.}\ \bibnamefont {Jian}},\ }\bibfield
  {title} {\enquote {\bibinfo {title} {Magic continuum in a twisted bilayer
  square lattice with staggered flux},}\ }\href {\doibase
  10.1103/PhysRevB.104.035136} {\bibfield  {journal} {\bibinfo  {journal}
  {Phys. Rev. B}\ }\textbf {\bibinfo {volume} {104}},\ \bibinfo {pages}
  {035136} (\bibinfo {year} {2021})}\BibitemShut {NoStop}%
\bibitem [{\citenamefont {Yu}\ \emph {et~al.}(2010)\citenamefont {Yu},
  \citenamefont {Zhang}, \citenamefont {Zhang}, \citenamefont {Zhang},
  \citenamefont {Dai}, ,\ and\ \citenamefont {Fang}}]{Yu2010}%
  \BibitemOpen
  \bibfield  {author} {\bibinfo {author} {\bibfnamefont {R.}~\bibnamefont
  {Yu}}, \bibinfo {author} {\bibfnamefont {W.}~\bibnamefont {Zhang}}, \bibinfo
  {author} {\bibfnamefont {H.-J.}\ \bibnamefont {Zhang}}, \bibinfo {author}
  {\bibfnamefont {S.-C.}\ \bibnamefont {Zhang}}, \bibinfo {author}
  {\bibfnamefont {X.}~\bibnamefont {Dai}}, , \ and\ \bibinfo {author}
  {\bibfnamefont {Z.}~\bibnamefont {Fang}},\ }\bibfield  {title} {\enquote
  {\bibinfo {title} {Quantized anomalous hall effect in magnetic topological
  insulators},}\ }\href {\doibase 10.1126/science.1187485} {\bibfield
  {journal} {\bibinfo  {journal} {Science}\ }\textbf {\bibinfo {volume}
  {329}},\ \bibinfo {pages} {61} (\bibinfo {year} {2010})}\BibitemShut
  {NoStop}%
\bibitem [{\citenamefont {Chang}\ \emph {et~al.}(2013)\citenamefont {Chang},
  \citenamefont {Zhang}, \citenamefont {Feng}, \citenamefont {Shen},
  \citenamefont {Zhang}, \citenamefont {Guo}, \citenamefont {Li}, \citenamefont
  {Ou}, \citenamefont {Wei}, \citenamefont {Wang}, \citenamefont {Ji},
  \citenamefont {Feng}, \citenamefont {Ji}, \citenamefont {Chen}, \citenamefont
  {Jia}, \citenamefont {Dai}, \citenamefont {Fang}, \citenamefont {Zhang},
  \citenamefont {He}, \citenamefont {Wang}, \citenamefont {Lu}, \citenamefont
  {Ma},\ and\ \citenamefont {Xue}}]{Chang2013}%
  \BibitemOpen
  \bibfield  {author} {\bibinfo {author} {\bibfnamefont {C.-Z.}\ \bibnamefont
  {Chang}}, \bibinfo {author} {\bibfnamefont {J.}~\bibnamefont {Zhang}},
  \bibinfo {author} {\bibfnamefont {X.}~\bibnamefont {Feng}}, \bibinfo {author}
  {\bibfnamefont {J.}~\bibnamefont {Shen}}, \bibinfo {author} {\bibfnamefont
  {Z.}~\bibnamefont {Zhang}}, \bibinfo {author} {\bibfnamefont
  {M.}~\bibnamefont {Guo}}, \bibinfo {author} {\bibfnamefont {K.}~\bibnamefont
  {Li}}, \bibinfo {author} {\bibfnamefont {Y.}~\bibnamefont {Ou}}, \bibinfo
  {author} {\bibfnamefont {P.}~\bibnamefont {Wei}}, \bibinfo {author}
  {\bibfnamefont {L.-L.}\ \bibnamefont {Wang}}, \bibinfo {author}
  {\bibfnamefont {Z.-Q.}\ \bibnamefont {Ji}}, \bibinfo {author} {\bibfnamefont
  {Y.}~\bibnamefont {Feng}}, \bibinfo {author} {\bibfnamefont {S.}~\bibnamefont
  {Ji}}, \bibinfo {author} {\bibfnamefont {X.}~\bibnamefont {Chen}}, \bibinfo
  {author} {\bibfnamefont {J.}~\bibnamefont {Jia}}, \bibinfo {author}
  {\bibfnamefont {X.}~\bibnamefont {Dai}}, \bibinfo {author} {\bibfnamefont
  {Z.}~\bibnamefont {Fang}}, \bibinfo {author} {\bibfnamefont {S.-C.}\
  \bibnamefont {Zhang}}, \bibinfo {author} {\bibfnamefont {K.}~\bibnamefont
  {He}}, \bibinfo {author} {\bibfnamefont {Y.}~\bibnamefont {Wang}}, \bibinfo
  {author} {\bibfnamefont {L.}~\bibnamefont {Lu}}, \bibinfo {author}
  {\bibfnamefont {X.-C.}\ \bibnamefont {Ma}}, \ and\ \bibinfo {author}
  {\bibfnamefont {Q.-K.}\ \bibnamefont {Xue}},\ }\bibfield  {title} {\enquote
  {\bibinfo {title} {Quantized anomalous hall effect in magnetic topological
  insulators},}\ }\href {\doibase 10.1126/science.1234414} {\bibfield
  {journal} {\bibinfo  {journal} {Science}\ }\textbf {\bibinfo {volume}
  {340}},\ \bibinfo {pages} {167} (\bibinfo {year} {2013})}\BibitemShut
  {NoStop}%
\bibitem [{\citenamefont {Otrokov}\ \emph {et~al.}(2017)\citenamefont
  {Otrokov}, \citenamefont {Menshchikova}, \citenamefont {Vergniory},
  \citenamefont {Rusinov}, \citenamefont {Vyazovskaya}, \citenamefont
  {Koroteev}, \citenamefont {Bihlmayer}, \citenamefont {Ernst}, \citenamefont
  {Echenique}, \citenamefont {Arnau},\ and\ \citenamefont
  {Chulkov}}]{Otrokov_2017}%
  \BibitemOpen
  \bibfield  {author} {\bibinfo {author} {\bibfnamefont {M.~M.}\ \bibnamefont
  {Otrokov}}, \bibinfo {author} {\bibfnamefont {T.~V.}\ \bibnamefont
  {Menshchikova}}, \bibinfo {author} {\bibfnamefont {M.~G.}\ \bibnamefont
  {Vergniory}}, \bibinfo {author} {\bibfnamefont {I.~P.}\ \bibnamefont
  {Rusinov}}, \bibinfo {author} {\bibfnamefont {A.~Yu}\ \bibnamefont
  {Vyazovskaya}}, \bibinfo {author} {\bibfnamefont {Y.~M.}\ \bibnamefont
  {Koroteev}}, \bibinfo {author} {\bibfnamefont {G.}~\bibnamefont {Bihlmayer}},
  \bibinfo {author} {\bibfnamefont {A.}~\bibnamefont {Ernst}}, \bibinfo
  {author} {\bibfnamefont {P.~M.}\ \bibnamefont {Echenique}}, \bibinfo {author}
  {\bibfnamefont {A.}~\bibnamefont {Arnau}}, \ and\ \bibinfo {author}
  {\bibfnamefont {E.~V.}\ \bibnamefont {Chulkov}},\ }\bibfield  {title}
  {\enquote {\bibinfo {title} {Highly-ordered wide bandgap materials for
  quantized anomalous hall and magnetoelectric effects},}\ }\href
  {https://doi.org/10.1088/2053-1583/aa6bec} {\bibfield  {journal} {\bibinfo
  {journal} {2D Mater.}\ }\textbf {\bibinfo {volume} {4}},\ \bibinfo {pages}
  {025082} (\bibinfo {year} {2017})}\BibitemShut {NoStop}%
\bibitem [{\citenamefont {Deng}\ \emph {et~al.}(2020)\citenamefont {Deng},
  \citenamefont {Yu}, \citenamefont {Shi}, \citenamefont {Guo}, \citenamefont
  {Wang}, \citenamefont {Chen},\ and\ \citenamefont {Zhang}}]{Deng2020}%
  \BibitemOpen
  \bibfield  {author} {\bibinfo {author} {\bibfnamefont {Y.}~\bibnamefont
  {Deng}}, \bibinfo {author} {\bibfnamefont {Y.}~\bibnamefont {Yu}}, \bibinfo
  {author} {\bibfnamefont {M.~Z.}\ \bibnamefont {Shi}}, \bibinfo {author}
  {\bibfnamefont {Z.}~\bibnamefont {Guo}}, \bibinfo {author} {\bibfnamefont
  {Z.~X.}\ \bibnamefont {Wang}}, \bibinfo {author} {\bibfnamefont {X.~H.}\
  \bibnamefont {Chen}}, \ and\ \bibinfo {author} {\bibfnamefont
  {Y.}~\bibnamefont {Zhang}},\ }\bibfield  {title} {\enquote {\bibinfo {title}
  {Quantum anomalous hall effect in intrinsic magnetic topological insulator
  mnbi2te4},}\ }\href {\doibase 10.1126/science.aax8156} {\bibfield  {journal}
  {\bibinfo  {journal} {Science}\ }\textbf {\bibinfo {volume} {367}},\ \bibinfo
  {pages} {895} (\bibinfo {year} {2020})}\BibitemShut {NoStop}%
\bibitem [{\citenamefont {Tateishi}\ and\ \citenamefont
  {Hirayama}()}]{Tateishi2021}%
  \BibitemOpen
  \bibfield  {author} {\bibinfo {author} {\bibfnamefont {I.}~\bibnamefont
  {Tateishi}}\ and\ \bibinfo {author} {\bibfnamefont {M.}~\bibnamefont
  {Hirayama}},\ }\bibfield  {title} {\enquote {\bibinfo {title} {Quantum spin
  hall effect from multi-scale band inversion in twisted bilayer
  ${\mathrm{bi}}_{2}{\mathrm{te}}_{1-x} {\mathrm{se}}_{x}$},}\ }\href@noop {}
  {\ }\Eprint {http://arxiv.org/abs/arXiv:2112.13770} {arXiv:2112.13770}
  \BibitemShut {NoStop}%
\bibitem [{\citenamefont {Balents}(2019)}]{Balents2019}%
  \BibitemOpen
  \bibfield  {author} {\bibinfo {author} {\bibfnamefont {L.}~\bibnamefont
  {Balents}},\ }\bibfield  {title} {\enquote {\bibinfo {title} {{General
  continuum model for twisted bilayer graphene and arbitrary smooth
  deformations}},}\ }\href {\doibase 10.21468/SciPostPhys.7.4.048} {\bibfield
  {journal} {\bibinfo  {journal} {SciPost Phys.}\ }\textbf {\bibinfo {volume}
  {7}},\ \bibinfo {pages} {48} (\bibinfo {year} {2019})}\BibitemShut {NoStop}%
\end{thebibliography}%


\end{document}